\documentclass[3p,12pt]{elsarticle}
\usepackage{ulem, url}

\usepackage{amssymb}

\usepackage[utf8]{inputenc}
\usepackage[dvipsnames]{xcolor}

\journal{Journal of Informetrics}

\begin{document}

\begin{frontmatter}

\title{Disturbance of questionable publishing to academia}

\author[ssu,pos1]{Taekho You}

\affiliation[ssu]{
            organization={School of AI Convergence, Soongsil University},
            addressline={369 Sangdo-ro}, 
            city={Dongjak-gu},
            postcode={06978}, 
            state={Seoul},
            country={Korea}}
\affiliation[pos1]{
            organization={Department of Industrial and Management Engineering, Pohang University of Science and Technology},
            addressline={77 Cheongam-ro}, 
            city={Pohang},
            postcode={37673}, 
            state={Gyeongsangbukdo},
            country={Korea}
            }
            
\author[kisti]{Jinseo Park}

\affiliation[kisti]{
            organization={Center for Global R\&D Data Analysis, Korea Institute of Science and Technology Information},
            adreessline={66 Heogi-ro},
            city={Dongdaemun-gu},
            postcode={02456},
            state={Seoul},
            country={Korea}
            }
            
\author[kisti]{June Young Lee}

\author[ssu]{Jinhyuk Yun\corref{cor1}}
\ead{jinhyuk.yun@ssu.ac.kr}
\cortext[cor1]{Corresponding authors}

\author[pos1,pos2]{Woo-Sung Jung\corref{cor1}}
\ead{wsjung@postech.ac.kr}

\affiliation[pos2]{
            organization={Department of Physics, Pohang University of Science and Technology},
            addressline={77 Cheongam-ro}, 
            city={Pohang},
            postcode={37673}, 
            state={Gyeongsangbukdo},
            country={Korea}
            }

\begin{abstract}
Questionable publications have been accused of ‘‘greedy’’ practices; however, their influence on academia has not been gauged. Here, we probe the impact of questionable publications through a systematic and comprehensive analysis with various participants from academia and compare the results with those of their unaccused counterparts using billions of citation records, including liaisons, \textit{i.e.}, journals and publishers, and prosumers, \textit{i.e.}, authors. Questionable publications attribute publisher-level self-citations to their journals while limiting journal-level self-citations; yet, conventional journal-level metrics are unable to detect these publisher-level self-citations. We propose a hybrid journal-publisher metric for detecting self-favouring citations among QJs from publishers. Additionally, we demonstrate that the questionable publications were less disruptive and influential than their counterparts. Our findings indicate an inflated citation impact of suspicious academic publishers. The findings provide a basis for actionable policy-making against questionable publications.
\end{abstract}


\end{frontmatter}

\section{Introduction}\label{sec:intro}
A market is a social system composed of stakeholders engaged in the exchange of commodities. From this perspective, academia is also a marketplace comprising scholars, journals and publishers, institutions, funding agencies, \textit{etc.} Based on the roles of these entities---most notably journals and scholars---academia can be classified into two broad business models: subscription-based and open access (OA). In both models, scholars have their research outcomes published in journals. The difference between the two models boils down to the party liable for article processing. The subscription-based model charges the readers, \textit{e.g.}, libraries, governments, institutions, and scholars, and limits the readership to a few solvent scholars and institutions. In contrast, the OA model allows everyone to access the full papers without payment; instead, they bill the authors an article processing charge (APC)~\cite{Bjork2012BMC, Eysenbach2005PLOSBio}. With the internet revolution, the OA model has gradually been accepted by academia. In the initial days of OA, scholars opined that the movement would enable them to give back to society, and consequently benefit academia and catalyse scientific literacy and advancement.

However, while this vision has partially materialised, a rise in the misuse of the OA model, which promotes publications of questionable standards with the sole purpose of profiteering, has been observed worldwide~\cite{Beall2012Nature, Darbyshire2018JCN, Memon2019JKMS}. Questionable journals (QJs) and questionable publishers (QPs) infiltrate academia with a perfunctory peer-review process by exploiting the weakness of the evaluation system, \textit{i.e.}, its reliance on journal indices. For instance, increasing academic competition has forced scholars to have more research published in indexed journals with high visibility, \textit{e.g.}, science citation index (expanded), social science citation index, and Scopus, \textit{etc}. QJs are often accidentally indexed in renowned citation databases despite the existence of preventive mechanisms. Journal indexing allows the expeditious publication of articles without peer review. This may entice authors who are at a critical juncture in their academic careers, increasing the likelihood of them preferring QJs. Critics of the OA model warned academia of the scientific stalemate and financial losses that would follow the publishing of unqualified articles~\cite{wiechert2019predator,dell2020trojan}. However, not all OA journals are questionable. Therefore, to protect the padawans from the dark side, critics have presented criteria to identify questionable publications along with a list of suspicious publishers and journals~\cite{Beall2012Nature, anderson2017cabell, kakamad2019kscien, Beall2016IHE}.

Despite the awareness of predatory publications, the market share of questionable publications has steadily increased (Fig.~\ref{fig:sm_questionable_size})~\cite{sterligov2016riding, machavcek2017predatory} as they carefully devise the stratagems necessary to expand their business in academia. QJs occasionally recruit respected editors to assert their credibility. Furthermore, they have also been found to appropriate the aliases of reputed scholars~\cite{Sorokowski2017Nature}. Authors pressed for time demand prompt article publication after submission ~\cite{Beall2016IHE, Shen2015BMCMedi, Cobey2019BMJ, Frandsen2019LeaPub, Shehata2018LeaPub}, which coincides with the offer by QJs. Recent studies informed the risks of questionable publishing, \textit{e.g.}, plagiarism~\cite{Owens2019JNS}, and insufficient peer review~\cite{Bohannon2013Science}. Notwithstanding the warnings from critics, the risks can also be deduced from qualitative evidence. However, this type of evidence is limited by the small sample of publications. Therefore, an unbiased and more quantitative approach is required to reveal the questionable publications. A few researchers have, in fact, quantitatively analysed the questionable publications, yet only gave limited insights on their behaviour~\cite{moussa2021citation,yan2011citation,kulczycki2021citation}. For instance, a study reported that papers published in QJs received fewer citations on average~\cite{moussa2021citation}; however, they only presented a simple citation count. Prior journal impact, which is often not considered in questionable publication studies~\cite{yan2011citation}, is a crucial factor determining if the paper should receive future citations. Citation distributions are known to be heavy-tailed, which implies that the majority of journals are rarely cited~\cite{waltman2012universality}. In other words, there are numerous journals besides the QJs that have been rarely cited. Thus, the difference between the QJs and the unaccused journals is insignificant when investigated at a superficial level as opposed to a fundamental one. The research question we attempt to answer in this study is: Do questionable publications sincerely harm the academic ecosystem?

The use of a journal self-citation as a proxy for spotting suspicious journals with inflated citation scores is a general procedure ~\cite{JCR2002, Fowler2007Sciento, Heneberg2016PONE, Wilhite2012Science}. Thus, quantifying a potential QJ's journal-level self-citation is a natural first step in achieving this. However, measuring self-citation may not be sufficient. A journal's strategy to decorate its citation score may not be directly reflected in the journal's self-citation, given the fact that most publishers publish multiple journals~\cite{Lariviere2015PONE,kojaku2021detecting}, leading authors to exchange citations between their journals without increasing journal-level self-citation score. As a result, publishers have taken over the role of independent journals in academic publishing. Therefore, instead of solely relying on self-citations, we must focus on the publishers who have not been considered owing to their reputation~\cite{eckert2018inside}.

In this study, we systematically analyse billions of citation records of questionable publications at various levels and of multiple agencies, \textit{i.e.}, papers, authors, journals, and publishers. With the large-scale data of 48,579,504 papers, 277,218 journals, and 2,714 publishers indexed and published in Scopus between 1996 and 2018~\cite{scopus}, we analysed the hidden citation patterns of questionable publications that can reflect graft. In addition to the simple citation count discussed in~\cite{moussa2021citation}, we also collected a set of journals closest to the QJs in terms of fame for comparison (see Materials and Methods). This approach allowed us to systematically monitor the unique patterns of questionable publications that distinguish them from the unquestioned publications.

A clear advantage of the comparative analysis of bibliographic data over mere investigations of the citation counts of QJs is that it allows one to compare the trajectories of the unquestioned journals (UJs) and QJs. That is, it does not overlook the possibility that the QJs are simply struggling because of the negative response of academia to their business model. However, the results of this analysis indicated that the QJs were not simply a struggling new model but were negatively affecting academia when compared to the reference UJs.

\section{Materials and Methods}
\subsection{Identifying questionable publications in Scopus}
\begin{figure}
    \centering
    \includegraphics[width=\textwidth]{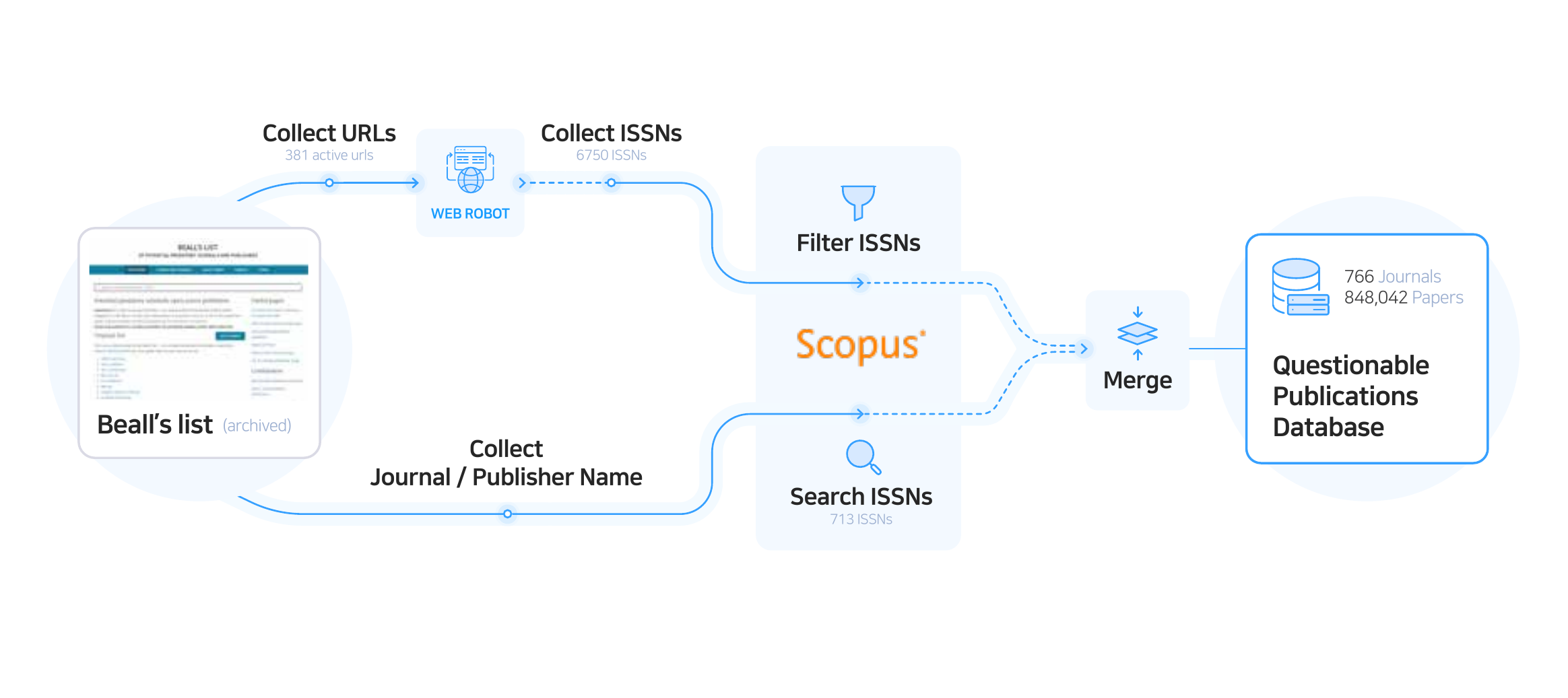}
    \caption{\textbf{Schematic diagram describing the steps of building the list of QJs from Scopus.} To improve credibility, we collect the journals by two methods and merge them: i) using a web crawler and ii) manual collection. In total, we collect 766 sources (journals) composed of 848,042 publications from Scopus.
    }
    \label{fig:process}
\end{figure}

We drafted a list of QJs from Beall’s list of potential predatory journals and publishers along with anonymous updates~\cite{BeallsWeb}. While Beall's list has known limitations, it is a widely adopted source of questionable publications~\cite{krawczyk2021open}. Our intention in choosing this list is not to criticize its constituents or contributors but merely to obtain a list of publishers and journals previously pointed out as potentially questionable. While the original Beall’s list was terminated in January 2017, it is still being updated by anonymous contributors (as of July 2021). We refer to Beall’s list as the one encompassing the original list as well as the one of the new website, unless specified otherwise. It contains two types of questionable publication lists: standalone journals and predatory publishers. We used both lists to identify questionable publications. To enhance the credibility of our list, we collected the data by two methods. First, we built a web robot to crawl the information on the journal web pages (Fig.~\ref{fig:process}). We programmed a bot with a Python 3.6 module using the Beautiful Soup package to extract the ISSN information from the websites of the journals and publishers in Beall’s list. The web crawling was performed from September 12 to October 12, 2018. Multiple accesses for the websites were attempted to bypass the effect of temporary outages of web services; however, the robot could only access 980 (including redirection to error pages) of 1,201 websites. We then collected 6,250 distinct ISSNs from the web pages of 381 journals or publishers. The rest of the websites do not contain ISSNs. For all subpages belonging to the target domain, we tested five consecutive items (5 grams) that appeared after the keyword \texttt{ISSN} or \texttt{ISSN:} separated by a space or punctuation, and found a 1-gram item formatted as \texttt{XXXX-XXXX}. We filtered out the ISSN candidates that failed to pass the ISSN checksum test~\cite{ISSNManual}.

In addition to web crawling, the journal and publisher websites were manually inspected to collect their names (Fig.~\ref{fig:process}). We accessed the websites in two periods in early 2019: from January 24 to February 28 for the original list, and from March 1 to 8 for the updated list. We used Beall’s list updated as of March 1, 2019, during which the connection was established on 953 websites among 1,236 websites. Subsequently, we searched the publisher and journal names on Scopus~\cite{scopus}, accompanied with Scimago Journal \& Country Ranking~\cite{sjr}, to obtain the ISSNs of the publications; this way, we collected 713 ISSNs of the journals in Scopus by manual inspection. Then we matched the ISSNs with the source list of Scopus to decode the \texttt{source id} of the QJs listed in Scopus. We selected the journals indexed in Scopus at least once between 1996 and 2018. Finally, we selected 766 sources comprising 848,042 publications from Scopus as the group of QJs.

\subsection{Selection criteria of the unquestioned journals}
Note that the expected number of citations for the QJs was significantly less than those in the entire Scopus database (Fig.~\ref{fig:sm_citation}). We believe the effect of reputation makes it unfair to compare the QJs to arbitrary journals. We chose the UJs that had a similar journal impact in the academic market as the QJs but that are not included in Beall’s list. Along with the journal impact, we also consider the journal’s subject classification and publication size. First, we calculated the journal impact by way of JCR impact factor~\cite{JCR2002}, but with Scopus data, as follows:

\begin{equation}\label{eq:impact}
\mbox{Journal Impact} = \frac{\mbox{citations received by papers published in past 2 years}}{\mbox{number of papers published in past 2 years}}.
\end{equation}

\noindent We should note that we consider journal impact as a proxy for two journals receiving similar attention in their field. For each subject category, denoted by its first two digits in the All Science Journal Classifications (ASJC), we arranged the journals in the order of their impact. We classified the journals into three groups by their publication volume to compensate for the journal size: large publishing (top 33rd percentile), moderate publishing (from 33rd to 66th percentiles), and small publishing (the rest; for details, see SI). Journals were considered inactive if they publish less than 30 papers annually and were excluded from the selection process. Subsequently, we chose the UJs by the following conditions: the journal i) was classified under the same subject category as the target QJ, ii) had the most similar journal impact with the target QJ, iii) and was in the same group of journal size. For a QJ classified under multiple subject categories, we separately considered one UJ per category. Accordingly, one QJ may pair with many UJs.

Another challenge is that the value of the citation impact may inflate owing to the increase in the overall number of publications. Thus, we normalised the citations~\cite{Peterson2019ResPol} by readjusting the number of citations for a given paper $i$ in the year $y$ as follows:

\begin{equation}
    C^*(y; i) = \frac{C(y; i)}{N_{top}(y)/N_{top}(2017)},
\end{equation}

\noindent where $C^*(y; i)$ is the normalised citation count for a given paper $i$ in the year $y$, $C(y; i)$ is the raw citation count for a given paper $i$ in the year $y$, and $N_{top}(y)$ is the number of articles in the top-cited field of study in the year $y$. The normalisation process made the journal impact independent of time, enabling the comparison of citations in papers published several years apart (see Fig.~\ref{fig:sm_IFnormalize}).

\subsection{Construction of the journal citation network}
To compute the centrality, we constructed a journal citation network for each year. We set the journals published in the target year as nodes and the citations received for two years after the target year as links. Note that the citation network was directed, and the in-degree was the same as the count of citations received for two years after publication. The out-degree corresponds to the total count of references published in the past two years. We demonstrated the robustness of the results for the network construction by validating them as follows: using references as links instead of citations, and using a different time window (for details, see SI).

\subsection{Z-score of atypical combination and disruptiveness index}
To measure the novelty of a study, we employed two well-known measures: the atypical combination of references~\cite{Uzzi2013Science} and disruptiveness index~\cite{Wu2019Nature}. First, we calculated the novelty by the atypical combination. We considered that a reference combination was atypical if the combination rarely appeared in the randomly shuffled citation network. We adopted the Markov Chain Monte Carlo method to generate the randomly shuffled citation networks, preserving the number of citations, the number of references, and the timeline of the citations. To handle the errors of random shuffling, we produced ten distinct random networks at the paper level. The z-score between two journals $i$ and $j$ was computed by standardising $z_{ij} = (o_{ij} - e_{ij})/\sigma_{ij}$, where $o_{ij}$ is the observed frequency of a journal pair in the original data, and $e_{ij}$ and $\sigma_{ij}$ are the average and standard deviations of frequency in the randomly shuffled network, respectively. A reference list in a single paper was composed of hundreds of pairs. To summarise the distribution of such z-scores, we examined the median and 10th percentile z-scores as representative statistics in terms of the central tendency and unusual combinations for a given paper.

We also calculated the disruptiveness index using the co-citation of publications~\cite{Wu2019Nature} as $D_x = (n_i - n_j)/(n_i + n_j + n_k)$, where $n_i$ is the number of publications that cite a publication $x$ while not citing any reference of $x$, $n_j$ is the number of publications that cite both $x$ and its references, and $n_k$ is the number of publications that cite any reference of $x$ but not $x$ itself. The disruptiveness index $D_x$ ranges from $(-1,1)$, where $D_x = -1$ indicates all publications citing both $x$ and its references (implying the least disruptive case), whereas $D_x = 1$ indicates all publications citing $x$ but none of its references (implying the most disruptive case).

\subsection{Disambiguation of scholars in Scopus}
Author disambiguation in Scopus is still imperfect. Therefore, we additionally merged the authors with a two-step disambiguation~\cite{Shultz2014EPJDS}. As the first step, we calculated the similarity between two papers using self-citation, the number of shared authors, the number of shared citations, and the number of shared references. If the similarity is higher than the threshold of $1$, the same author wrote the two papers. We merged the groups when the average similarity between two groups was beyond the threshold of $0.19$. To this end, we used the parameters from the original study~\cite{Shultz2014EPJDS}. The disambiguation calculations for the uncited single-authored papers may encounter errors. In this case, the author was classified as a distinct author who published only one article; we excluded these authors and their papers after the disambiguation to eliminate the errors because the volume of their papers accounted for less than 1\%. We used the merged authors to calculate the author-based metrics. For instance, the authors’ academic age was computed as the difference between her/his first and last publications. Before disambiguation, there were 21,805,486 Scopus author ids (\texttt{auid}) with 1,895,356 unique author names, which were merged into 15,022,380 unique author names after the disambiguation.

\section{Results}\label{sec:results}
The main objective of this study was to detect anomalies in the citation patterns of QJs. We used the January 2019 dump of Scopus CUSTOM XML DATA, which includes all Scopus citation records from 1996 to 2018, along with the bibliographic metadata. Because the dataset itself does not reveal QJs, we identified them by systematically extracting the ISSNs from the journal websites using Beall’s list of potential predatory journals and publishers (see Materials and Methods). To improve the accuracy, we manually cross-verified the QJs by collecting the names of their journals and publishers. With the acquired ISSNs and titles, we identified the \texttt{Scopus source id} for each journal. In total, we collected 766 QJs indexed in Scopus at least once between 1996 and 2018. To proceed to the comparative analysis of QJs, we emphasise the fact that a journal’s reputation strongly influences the chance of its future citation~\cite{yan2011citation}. The effect of the accumulated citation count itself makes it infeasible to directly compare the citation patterns of the journals with different degrees of reputation. To compensate for the effect of reputation, we collected the reference journals using the journal impact, which is estimated by way of the Journal Citation Reports (JCR)~\cite{garfield2006history}. The criteria for selecting the reference journals were as follows: the reference journals i) shall not be present in the list of QJs, ii) shall be in the same subject category as the target QJ, and iii) shall have a journal impact the most similar with that of the target QJ (see Materials and Methods). The reference journals are the same as the UJs mentioned above unless specified otherwise. Using these two groups, we performed a comparative investigation of the citation patterns from various aspects.

\subsection{Publishers and journals as liaisons in academia}\label{sec:publishers}

\begin{figure}
    \centering
    \includegraphics[width=\textwidth]{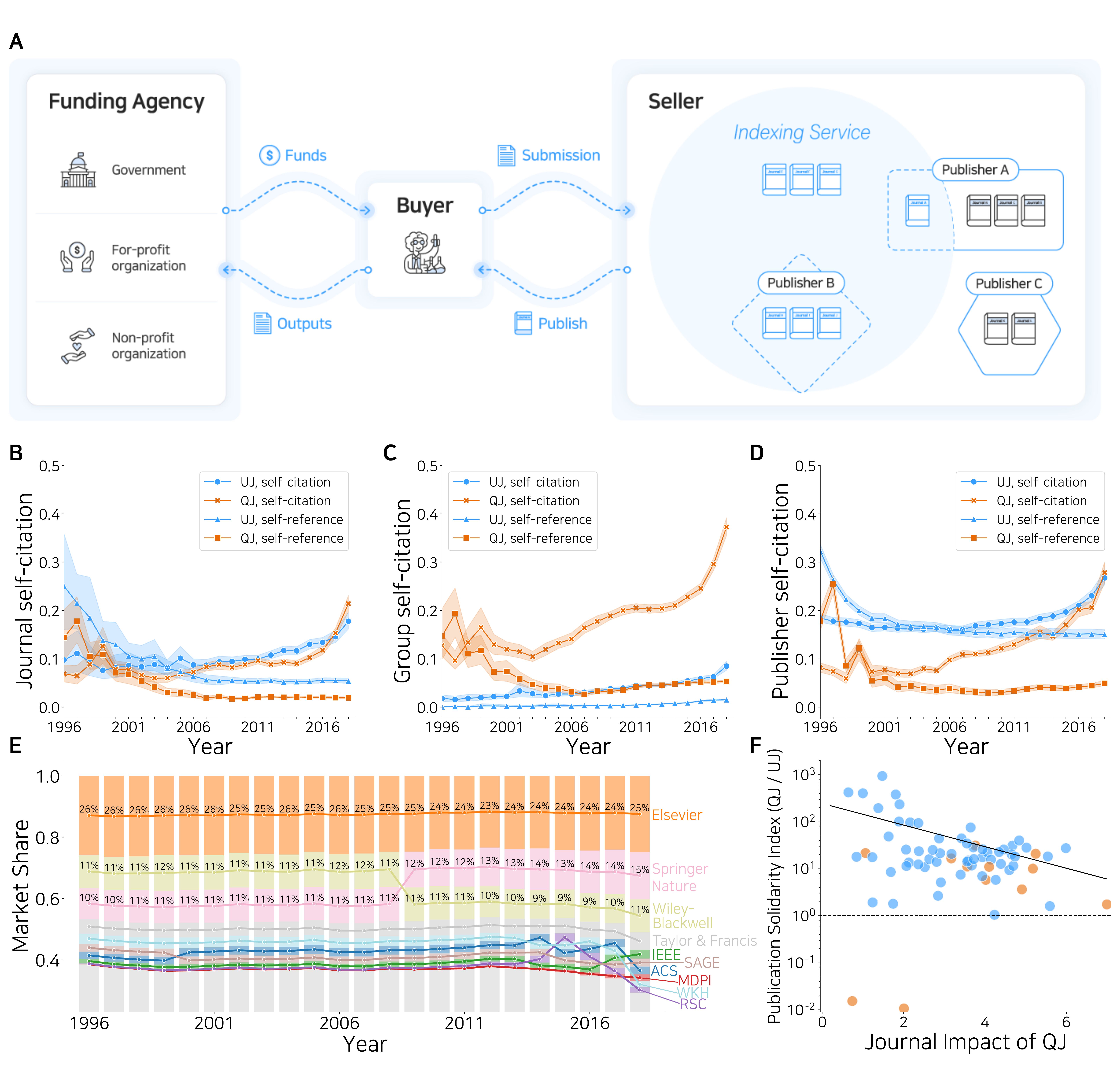}
    \caption{\textbf{Publication patterns as citation sources and publication references in the two groups.}
    (\textbf{A}) Schematic diagram describing the composition of the academic marketplace from an economics perspective. Funding agencies investigate the researchers. Researchers submit their research works to a journal to purchase pages in the journal (as buyers). Journals and publishers sell pages in the journal to the researcher (as vendors). The researchers pay an APC or a subscription fee while providing the results to the funding agencies.
    (\textbf{B})--(\textbf{D}) Mean self-citation and self-reference rates in three different levels of QJs (orange lines) and UJs (blue lines). The rates were calculated as the ratios of self-citations (self-references) from the identical journal (\textbf{B}), category (QJs and UJs, respectively, \textbf{C}), and the publisher (\textbf{D}). The shaded areas represent the $95\%$ confidence intervals.
    (\textbf{E}) Market shares of the top ten publishers in terms of the number of publications, defined as the ratio of the number of articles for a certain publisher to the total number of all publishers. Note that the market share value of the top three publishers is dominant, which is approximately $50\%$
    (\textbf{F}) Ratio of publication solidarity index $\psi$ of the corresponding QJ to that of the UJ. The dotted guide line represents the null model of equal $\psi$ for the QJ and UJ. The colour represents the relative size of the publishers: the orange dots indicate that the QP is larger than the UP, whereas the blue dots indicate that the QP is smaller than the UP. The solid regression line displays the fitted model of $y \sim e^{-0.53x}$, where $x$ is the journal impact of QJ.
    }
    \label{fig:provider}
\end{figure}

As mentioned earlier, a market is composed of buyers and sellers. In this sense, academia can also be considered a market sustained by the exchange of knowledge between authors (buyers) and journals/publishers~(vendors; see Fig.~\ref{fig:provider}A). Therefore, both parties must be accounted for to comprehend the entire landscape of academia. We began with the publishers and journals, which are the liaisons of academia. Journals select the articles to be distributed in academia based on their scope; thus, the direction of a journal is reflected in their publication record. Additionally, the citation pattern also reflects their strategy because citations are widely used to gauge the performance or reputation of a journal~\cite{janosov2020success,moed2006citation}. For instance, the journal’s self-citation rate indicates the veracity with which it pursues high citation scores; however, indexing services have been restricting this endeavour. Similarly, we introduced two simple metrics to characterise the collective citation patterns: the citation rate $R_c(i;j)$ and reference rate $R_r(i;j)$, to account for certain journals (publishers) $i$ and $j$ as follows.

\begin{equation} \label{eq:Rc}
R_{c}(i; j) = \frac{C_{ji}}{\sum_{a}{C_{ai}}}, 
\end{equation}
\begin{equation} \label{eq:Rr}
R_{r}(i; j) = \frac{C_{ij}}{\sum_{a}{C_{ia}}},
\end{equation}

\noindent where $C_{ij}$ is the number of citations a journal (publisher) $i$ attributes to a paper published in the target group $j$, whereas $a$ is the set of all journals in the dataset. Here, the target group $j$ can be one of the following: i) an individual journal, ii) a group of journals, iii) a publisher composed of multiple journals, or iv) a group of publishers (Fig.~\ref{fig:provider}B--D). The value of this metric increases when the collective volume of citations has increased for a certain group; thus, this value represents the degree of a citation preference between the groups. Hereafter, for the special case of $i=j$, metrics $R_{c}(i; i)$ and $R_{r}(i; i)$ shall be referred to as the self-citation and self-reference rates, respectively. To illustrate with an example, a publisher’s self-reference rate can be calculated by $\sum_{j\in P} R_r(i;j)$, where $P$ is a set of journals belonging to the same publisher of journal $i$; that is, the fraction between the total count of references citing the publisher and the total number of references to a journal.

The logical step forward is to investigate the difference between QJs and UJs using these metrics. We pay special attention to the distinct patterns of the QJs, and therefore examine the difference between the mean self-reference rates of the QJs and UJs. We observed that the self-reference rate of the QJs was less than that of the UJs (Fig.~\ref{fig:provider}B). The mean self-citation rate of the QJs was also less than that of UJs, except in 2017 and 2018. In summary, the journal-based self-citation metric did not reveal any peculiar behaviour from the QJs.

On the contrary, we observed that the QJs had a collective tendency to preferentially cite other QJs (Fig.~\ref{fig:provider}C). Specifically, the QJs received ${\sim}20\%$ of the citations from other QJs, and only ${\sim}7\%$ from the UJs (Fig.~\ref{fig:provider}C). Note that we selected UJs from the same list of journals as the QJ in terms of journal impact, size, and subject category (see Materials and  Methods). Thus, the QJs and UJs were expected to be in an equivalent proportion. However, more QJs have received citations from other QJs than UJs have received from other UJs. In other words, the journal impact of QJs depends more on citations from other QJs than from UJs. This observation along with the lower self-citation rate of QJs indicates a possible citation preference among the publishers of QJs.

The contrasting observation raises an intriguing question: Does the citation preference indeed originate from the publisher? To answer this question, we inspected the self-citation preference of the publishers (Fig.~\ref{fig:provider}D). By calculating the fraction of the publishers’ self-citation for both the QJs and UJs, we found that the UJs have cited the same publisher more times than the QJs. However, the heterogeneity in the number of publications by publishers possibly affected the self-citation and self-reference probabilities. That is, it is normal to cite more studies from a larger, more reputed publisher despite the lack of any explicit preference. We also found a large heterogeneity in the size of publishers (see Fig.~\ref{fig:sm_publisher_dist})~\cite{Lariviere2015PONE}. Large publishers, such as Springer Nature, Elsevier, Wiley-Blackwell, and Taylor \& Francis, have increased their market share and account for half of all academic publications (Fig.~\ref{fig:provider}E). This stark heterogeneity makes it unfair to directly compare the publisher self-reference rates. Therefore, we must account for the market share of the publishers. To compensate for the heterogeneity, we propose the publication solidarity index $\psi(i)$ for a journal $i$, which is expressed as follows:

\begin{equation}\label{eq:psi}
\psi(i) = \frac{1}{\sum_{j|j \in P_i} N_j} \cdot \frac{\sum_{j|j \in P_i}{R_r(i;j)} / Q_r}{\sum_{j|j \in P_i}{R_c(i;j)} / Q_c},
\end{equation}

\noindent where $Q_r$ and $Q_c$ are the expectations of self-reference and self-citation rates for a publisher, respectively, which are defined as:

\begin{equation}
Q_r = \frac{\sum_{k|k \in P_i} \sum_{j|j \in P_i} C_{kj}}{ \sum_{k|k \in P_i} \sum_j C_{kj}},
\end{equation}

\begin{equation}
Q_c = \frac{\sum_{k|k \in P_i} \sum_{j|j \in P_i} C_{jk}} {\sum_{k|k \in P_i} \sum_j C_{jk}}, 
\end{equation}

\noindent and $N_j$ is the total number of papers published in journal $j$. The term $C_{ij}$ is the number of citations received by journal $j$ from papers in journal $i$, and $P_i$ is the set of journals in the publisher of journal $i$. 

The self-reference rate shows the number of times the journal cites its publisher, which cannot adequately represent the entire academic landscape. However, a high self-reference rate is not necessarily evidence of malpractice. First, as shown in Fig.~\ref{fig:provider}D, large publishers tend to cite more (see Fig.~\ref{fig:sm_publisher_cite}). Additionally, the self-reference rate can be high if there is no other journal covering the same subject. For instance, a high-profile journal may publish several significant studies in quick succession, and therefore researchers have few options from which to cite. Occasionally, it unavoidably cites papers published in the most reputed journals~\cite{Lariviere2015PONE,Zhou21Sciento}. Here, $\psi$ was designed to compensate these biases: we divided the self-reference rate i) by the size of the publisher as ${1}/{\sum_{j|j \in P_i} N_j}$ to compensate for the size effect, and ii) by the self-citation rate as $\sum_{j|j \in P_i}{R_c(i;j)} / Q_c$ to account for the indispensability of a citation. Therefore, the proposed metric is mostly free from the influence of the publisher size and journal’s profile. It has a high value when the journal assigns more citations to their publisher than the citations the latter receives from other sources (see~\ref{sec:sm_psi_validation}).

To characterise and distinguish the citation behaviour of the QJs from that of the UJs, we compared the $\psi(i)$ of QJs and their corresponding UJs. Note that we excluded standalone QJs and UJs because we designed the metric to capture the self-reference rates of publishers composed of multiple journals. We then calculated the ratio of $\psi$ of the QJs to their corresponding UJs; a ratio of $>1$ implies the QJs have excessively self-referenced, and a ratio of $<1$ indicates \textit{vice versa}. We found that, despite two exceptions, the QJs have a higher $\psi$ than the UJs (Fig.~\ref{fig:provider}F). The $\psi$ value may largely depend on the publisher size; however, the result is robust to the relative publisher size between QJs and UJs. The QJs have larger $\psi$ values even if the QP is larger than the unquestioned publisher (UP).

The volume of the QP is observed to be larger than that of the UP for the two exceptions, where the ratio was $\psi < 1$. To verify this, we manually examined the two cases. The first exception is the \textit{Journal of Essential Oil-Bearing Plants}, which was acquired by an UP in 2013; subsequently, the journal’s metric and $\psi$ improved. Specifically, CiteScore increased from 0.5 to 1.3, and $\psi$ from 0.1 to 0.01 between 2013 and 2018. The publisher of the second journal, \textit{Insects}, was eventually excluded from Beall’s list. In summary, QJs are likely to embellish their journal impact with self-reference. Note that even though the ratio of $\psi(i)$ is higher than 1 for all QJs (excluding the two exceptions), the ratio is decreasing with the journal impact of the QJs (solid line in Fig.~\ref{fig:provider}F). Thus, once the QJs have gained a place in academia, they gradually reduce publisher self-references.

\subsection{Authors as prosumers in academia}\label{sec:authors}
Thus far, we investigated the role of publishers and journals, which are the liaisons in academia. The logical step forward is to observe the response of authors to QJs. Despite several warnings, the number of unique authors publishing in QJs has increased over time (Fig.~\ref{fig:sm_author_size}). Note that the author decides whether to submit a paper. If submitted, there are two possibilities: the authors may submit an article to QJs with or without prior knowledge. Thus, to inspect the author’s behaviour, we must answer the question, ‘‘Can authors distinguish QJs from journals with similar external appearances?’’ Moreover, we must investigate any unfair advantage they received from the questionable publication to fully understand their behaviour.

\begin{figure}
    \centering
    \includegraphics[width=0.7\textwidth]{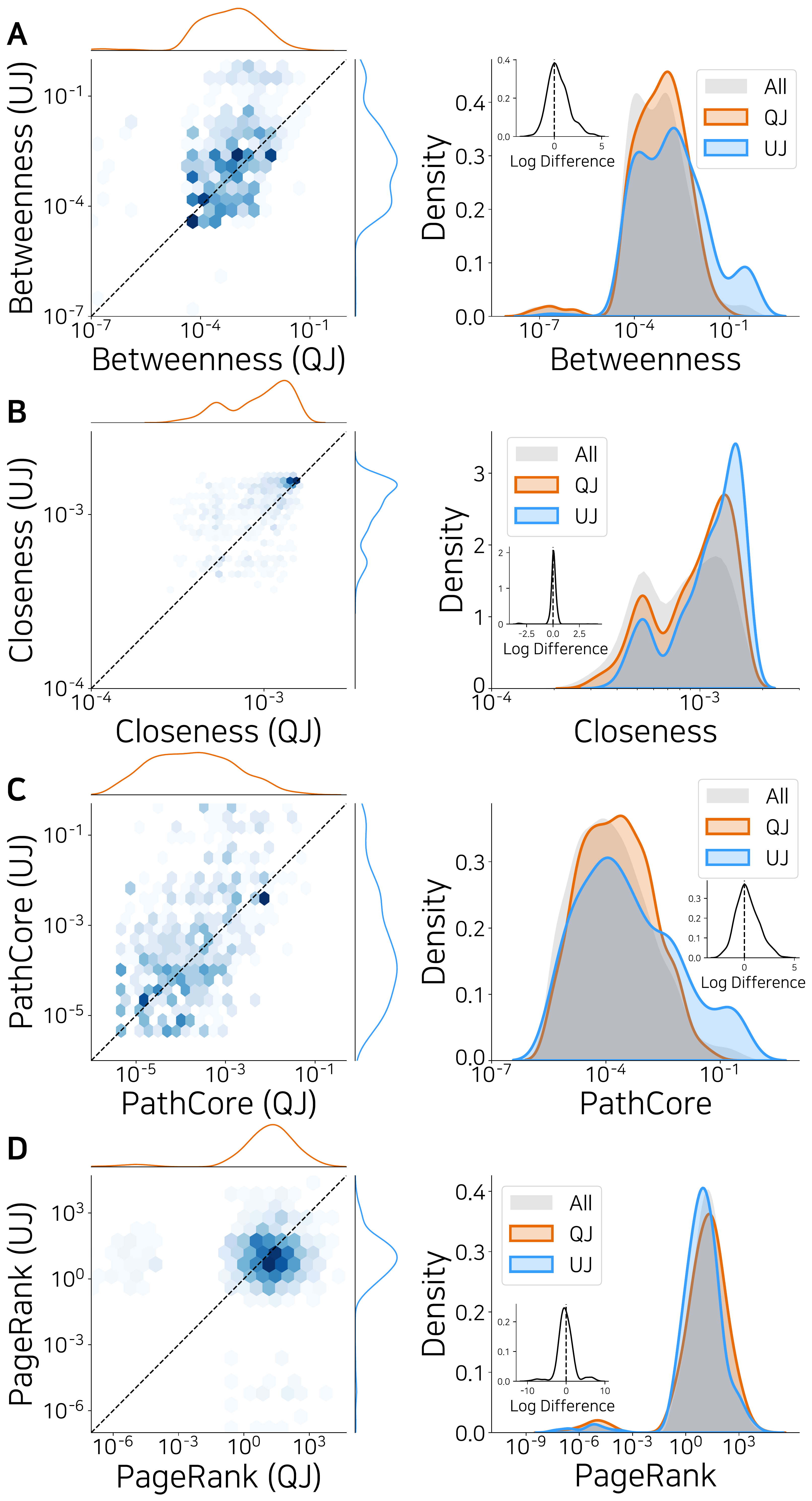}
    \caption{\textbf{Centrality distribution of QJs and UJs in the journal citation network}: (\textbf{A}) Betweenness (BC), (\textbf{B}) closeness (CC), (\textbf{C}) PathCore (PC), and (\textbf{D}) PageRank (PR) centralities. The centralities for the journal citation network are calculated using the two-years received citations of publications from 2016. The leftward heatmaps show the interrelation between the centralities of a QJ and those of the corresponding UJ. The rightward distributions display the kernel density estimation results of centrality distribution for QJs (orange), UJs (blue), and all journals (grey), while the insets present the log difference of each centrality score between UJs and QJs.
    }
    \label{fig:journal}
\end{figure}

To understand the authors’ response to the rise of questionable publications, we first consider their ability to discriminate the questionable publications from the rest, especially from the unquestioned publications. According to the calculations in Section \ref{sec:publishers}, the QJs demonstrated a higher $\psi$ than the UJs, which indicates the homophilic citations of QJs. Because we selected UJs with similar citation impacts and publication sizes, the above results demonstrated highly clustered citations for the QJs that are possibly located in the periphery of the academic market because they are rarely cited (Fig.~\ref{fig:sm_citation}). To prove this hypothesis, we constructed a journal citation network consisting of all the Scopus (see Materials and Methods) journals (as nodes) and citations (as links). We measured various network centralities, namely betweenness centrality (BC), closeness centrality (CC), PathCore score (PC), and PageRank centrality (PR). The important journals in the network, with a constructed network, have large centralities~\cite{Freeman1979Social,Cucuringu:2016dk,page1999pagerank,gonzalez2010centrality}, meanwhile, each centrality captures a different role of the node. Journals with a high BC score connect disciplines, while journals with a high CC score are in the center of the network. The core-periphery structure can be captured by PC as a probability that a journal belongs to the network's core. A high PR score indicates that the journals are influential in the network.

According to the results, the centrality distributions of the UJs are likely to have a heavier tail than those of the QJs, except PageRank (Fig.~\ref{fig:journal}, Fig.~\ref{fig:sm_centrality_highjif}, Fig.~\ref{fig:sm_centrality_lowjif}). Although QJs tend to be distributed over a higher PageRank range than UJs (Fig.~\ref{fig:journal}D), this does not imply that QJs are more important than UJs in the academic market. For instance, with an increment in self-loops, the number of nodes with a high PageRank also increases~\cite{Ghasemieh2016influence}. As discussed in the previous section, we found that QJs show a higher $\psi$, which underscores the highly clustered cycles in the citation network. The trend is also valid for directly comparing the centralities of QJs and UJs. Specifically, $61.6\%$ UJs exhibited high BC, $69.7\%$ high CC, and $56.0\%$ high PC. Only $43.8\%$ UJs showed a higher PageRank than the QJs. Note that this observation is robust for the size of the citation window, target year, and/or the link types (see Materials and Methods and Figs.~\ref{fig:sm_BC2y}--\ref{fig:sm_PC5y_ref}). Overall, despite having a similar discipline, journal size, and citation impact, UJs served more central roles than QJs in academia.

\begin{figure}
    \centering
    \includegraphics[width=\textwidth]{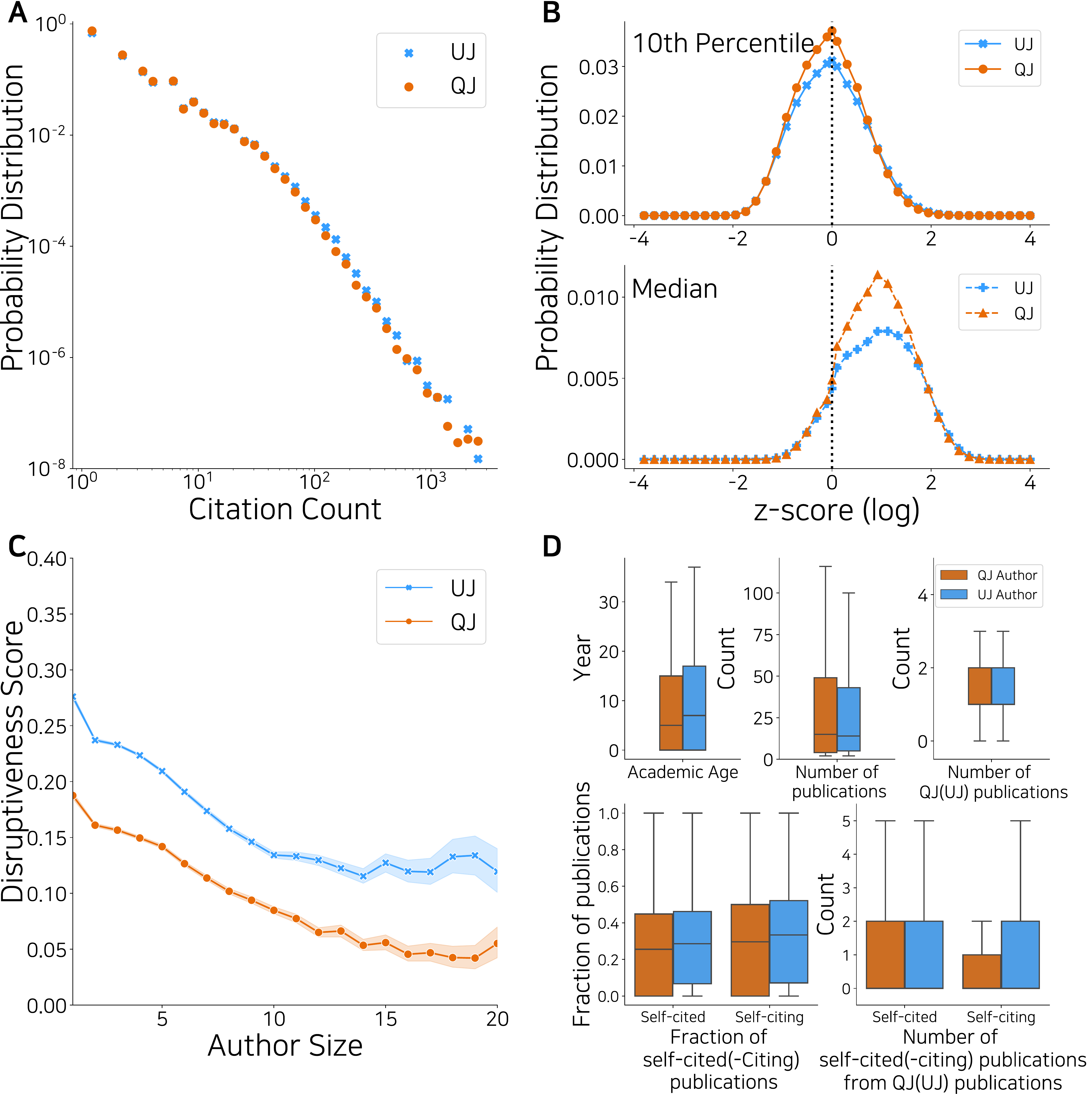}
    \caption{\textbf{Characteristics of papers and demographics of authors in QJs and UJs}.
    (\textbf{A}) Citation distribution for QJs and UJs.
    (\textbf{B}) Probability density of atypical combinations on the (upper) 10th percentile and (lower) median~\cite{Uzzi2013Science}. The z-scores are computed for journals in the references pairs and are found to have low values when the combination is rarely co-referenced in other papers.
    (\textbf{C}) Average disruptiveness index~\cite{Wu2019Nature} for papers by the number of authors. The index is low if the paper has several shared references with subsequent papers, which indicates that it details a follow-up study repeating the precedents. A previous study showed that the disruptiveness index had a negative correlation with the number of authors. Therefore, we display the index for a given number of authors.
    (\textbf{D}) Demographics of authors publishing papers in QJs and UJs and citation statistics. Each subpanel displays: (upper-left) the academic age between their first and last articles, (upper-mid) the number of papers, (upper-right) the number of papers in QJs and UJs, (lower left) the fraction of author-level self-cited and self-citing articles in their careers, and (lower right) the number of group-level self-cited and self-citing papers, belonging to UJs or QJs. On average, QJ authors are 8.09 academic years old with 50.6 articles, 2.11 QJ articles, $27.1\%$ ($31.2\%$) author-level self-cited (-citing) papers, and 2.12 (1.03) group-level self-cited (self-citing) articles. Meanwhile, UJ authors are 8.94 academic years old with 41.9 articles, 2.67 UJ articles, $28.5\%$ ($32.7\%$) author-level self-cited (self-citing) papers, and 2.35 (2.38) group-level self-cited (self-citing) articles.
    }
    \label{fig:indi}
\end{figure}

To further understand the authors’ response to questionable publications, we examined their unfair advantage from questionable publications. The benefit of studying the massive citation dataset is that the citation pattern of individual authors and publications can be identified. The advantage of having a paper published in a QJ is the fast processing time. However, this advantage is accompanied by inadequate peer review, risking the publishing of underqualified research~\cite{Cobey2019BMJ,Bohannon2013Science}. Because we selected UJs with similar citation impacts and sizes as their questionable counterparts, their citation distributions are similarly heavy-tailed (Fig.~\ref{fig:indi}A). We also found that QJs publish more uncited articles (27.5\%) than UJs (22.2\%), albeit by an insignificant margin. In terms of the citation distribution itself, the two groups are nearly identical. Therefore, any differences in the quality of articles may result from surplus citations.

However, simple citation counts are not always an accurate measure of the quality, \textit{e.g.}, the baseline of citation measures are largely determined by the discipline~\cite{lillquist2010discipline}. Therefore, we must account for the hidden context of citations, which better reflects the quality of a publication. For instance, novelty plays an important role in the progress of science and technology and is measured using atypical combinations of references~\cite{Uzzi2013Science}. Following this idea, we computed the z-score, which represents the rarity of reference pairs compared to the random null model (see Materials and Methods). A z-score of less than zero indicates that the reference combination is relatively rare and novel. On average, a single paper has more than ten references, which are composed of hundreds of pairs. Therefore, we selected the 10th percentile and median of the z-scores as the representative statistics. The majority of the 10th percentile z-scores are located near zero for both UJs and QJs, with no significant difference between them (see Fig.~\ref{fig:indi}B). There is also nearly zero disparity in the median z-score. In summary, both QJs and UJs have articles of a similar degree of novelty on average.

It may be argued that the novelty of the reference pairs cannot capture the type of contribution to academia because it does not consider the future citation impact. However, highly disruptive articles have the potential to introduce breakthroughs in science and technology, and thus have a large impact on society. From this perspective, there are two types of contributions to science and technology: i) disruptive research, which introduces new approaches and fundamental questions in the field, and ii) developmental research, which aims to practice well-known theories in new contexts~\cite{Wu2019Nature}. We adopted the disruptiveness index to measure the degree of disruption for individual papers, accounting for both references and citations simultaneously (see Materials and Methods)~\cite{Wu2019Nature}. We demonstrated that QJs tend to publish fewer disrupting research papers. Disruptiveness decreases with the number of coauthors who contributed to the original study~\cite{Wu2019Nature}. The disruptiveness index of QJs is consistently less than that of UJs (see Fig.~\ref{fig:indi}C). It saturates at $\sim0.14$ for UJs, and at $\sim0.05$ for QJs, if more than 15 persons co-authored the paper. When combined, despite having a similar citation impact, the manuscripts published in QJs are more conventional than those of UJs.

Additionally, we examined the demographics of authors in terms of their publication preference (Fig.~\ref{fig:indi}D). On average, QJ authors are relatively younger in terms of academic experience than those published in UJs (8.09 vs. 8.94 years, respectively). On the one hand, QJ authors publish more in their careers (8.7 publications more than the UJ authors on average), but QJ authors have 2.11 questionable publications while UJ authors have 2.67 unquestioned publications. Thus, QJ authors are more productive than UJ authors; however, QJ authors have less questionable publications while UJ authors have more unquestioned publications.
On the other hand, QJ authors had $27.1\%$ self-citations in their manuscripts, while also having been self-cited by $31.2\%$ of their publications. This is a marginally smaller proportion compared with the UJ authors, who have $28.5\%$ self-citations in their manuscripts and were self-cited by $32.7\%$ of their publications. QJ and UJ authors have a similar number of publications cited by QJs and UJs, i.e., 2.12 and 2.35, respectively. On average, QJ authors cited questionable publications only once in their career, while UJ authors cited unquestioned publications twice on average.

To summarise, authors tend to have less important research published in QJs, owing to inadequate peer review. QJs are located on the fringes of academia. Papers published in QJs are not as disruptive as those in UJs; however, they receive the same number of citations on average. Although the motivation of each author cannot be pinned, the above findings led to the conclusion that authors who consistently publish in QJs generally anticipate exploitation for profiteering by the publishers.

\section{Discussion}
In this study, we analysed the massive bibliographic metadata of Scopus to inspect the status quo of questionable publications as distinguished from unquestioned publications. Previous studies have investigated QJs in academia through small samples or simple citation counts, which have raised many unanswered questions~\cite{Beall2012Nature,Owens2019JNS,jain2019evolving}. We attempted to quantitatively understand the behaviour of questionable publications. The analysis of journal self-citation and reference revealed self-favouring citations among QJs from publishers. Moreover, QJs adopt a less important role in the journal citation network, as reflected in the centrality metrics. Authors had few creative research works published in the QJs, as highlighted by the disruptiveness index. Observational data suggest that a comprehensive analysis of citation patterns could support the case against QJs, which will help dispute deterioration in academia. The limitation of this study is the diversity of the sample; it only considers the journals indexed in Scopus, which do not cover all dynamics on QJs. Thus, the self-favourable behaviour (or any other singular behaviours) of non-indexed journals should be considered in a future study. Moreover, the indexed QJs can be thought of as they are approved by the existing community, yet it still needs to be validated~\cite{bagues2019walk}. In addition, one should note that Beall's list has been criticised for its selection criteria~\cite{krawczyk2021open}. There are alternative lists, such as Cabells' Predatory Reports, yet only a small number of journals on this list are indexed in the Scopus or the Web of Science. Thus, it is hard to collect their citation history. If one encompasses the citation patterns outside the Scopus and the Web of Science, it may yield a more accurate status of questionable publications; this is left for further study. The sample and a small subset of all questionable publications qualify as questionable publications according to the criteria set by indexing services~\cite{ross2019predatory, duc2020predatory}, which may contain a potential bias of data. Thus, investigation of non-indexed journals using broader datasets (for example, web-collected datasets such as Google Scholar or Microsoft Academic Graph) may provide additional insights, which we have left for future study. However, non-indexed journals have insignificant visibility, and therefore, their influence on academia is limited. Another limitation is that the criteria used to select UJs is imperfect, as the list of UJs might contain hidden QJs. We did not consider the fluctuations in citation patterns due to individual authors and publications because we investigated the collective tendencies of several publications. We also omitted possible self-favouring groups in UJs for a future study. 

Governments, institutions, and funding agencies occasionally promote researchers by incentivising their research outcomes~\cite{quan2017publish,kim2016scientists}. Evaluations of such incentives conventionally rely on quantitative measures, \textit{e.g.}, the number of publications, the number of citations, and the reputation (citation impact) of published journals. However, this study suggests that the comprehensive quality of research outcomes should also be assessed on the value of its contents, and not solely computed through simple quantitative metrics. In addition, recent reports suggest that authors in low-income countries are more likely to avail the services of questionable publications. This warrants the active consideration of the inequality of science among countries~\cite{lalu2017stakeholders,demir2018predatory}. For instance, the publication fees of major journals exceed the annual income of certain countries; the current APC waiver policy does not solve this problem. Thus, one could argue that the emergence of QJs is a natural market reaction to increased competition. Indeed, a survey has demonstrated that many researchers from developing countries believe they will be rejected by Western journals and therefore seek publication in alternative journals (that is, QJs)~\cite{kurt2018authors}. Simultaneously, they believe they lack adequate training in research methodology and reporting, and thus are unable to submit to traditional journals~\cite{kurt2018authors}. Therefore, QJs may represent a new market response to the somewhat rigid academia of today. However, our findings, combined with insights of previous studies, indicate that QJs are not yet a part of a healthy academic environment~\cite{Sorokowski2017Nature,Shen2015BMCMedi,bagues2019walk}. Despite this, we sincerely hope that QJs will become more wholesome members of academia as a result of future changes in direction. We believe each stakeholder should have a role in cultivating academia. A suitable policy, accompanied by quantitative analysis, judiciously combined with context-specific assessment, will help introduce innovative ideas into academia worldwide.

\section*{Authors' contribution}
All authors designed the experiment and wrote the manuscript, TY, JP, JY collected the data. TY analysed the data. All authors reviewed, edited, and approved the final draft.

\section*{Acknowledgements}
The National Research Foundation (NRF) of Korea Grant funded by the Korean Government supported this work through Grant No. NRF-2020R1A2C1100489 (T.Y. and J.Y.) and NRF-2021R1F1A1063030 (W.S.J.). The Korea Institute of Science and Technology Information (KISTI) also offered institutional support for this work (K-21-L03-C07-S01; J.P. and J.Y.L.) and provided KREONET, the high-speed internet connection. The funders had no role in the study design, data collection and analysis, decision to publish, or preparation of the manuscript.

\bibliographystyle{elsarticle-num} 
\bibliography{ref_questionable}

\clearpage
\begin{center}
    \item {\fontsize{14}{0}\selectfont \textbf{Supplemental Information for}}
    \item{\fontsize{14}{0}\selectfont Disturbance of questionable publishing to academia}
    \item{\fontsize{10}{0}\selectfont Taekho You, Jinseo Park, June Young Lee, Jinhyuk Yun$^{*}$, Woo-Sung Jung$^{*}$}
    \item{\fontsize{10}{0}\selectfont $^*$Corresponding author. Email: jinhyuk.yun@ssu.ac.kr; wsjung@postech.ac.kr}
\end{center}

\setcounter{equation}{0}
\setcounter{figure}{0}
\setcounter{table}{0}
\setcounter{page}{1}
\setcounter{section}{0}

\makeatletter
\renewcommand{\thesection}{Section S\arabic{section}}
\renewcommand{\theequation}{S\arabic{equation}}
\renewcommand{\thefigure}{S\arabic{figure}}
\renewcommand{\figurename}{\textbf{Fig.}}

\begin{figure}
    \centering
    \includegraphics[width=\textwidth]{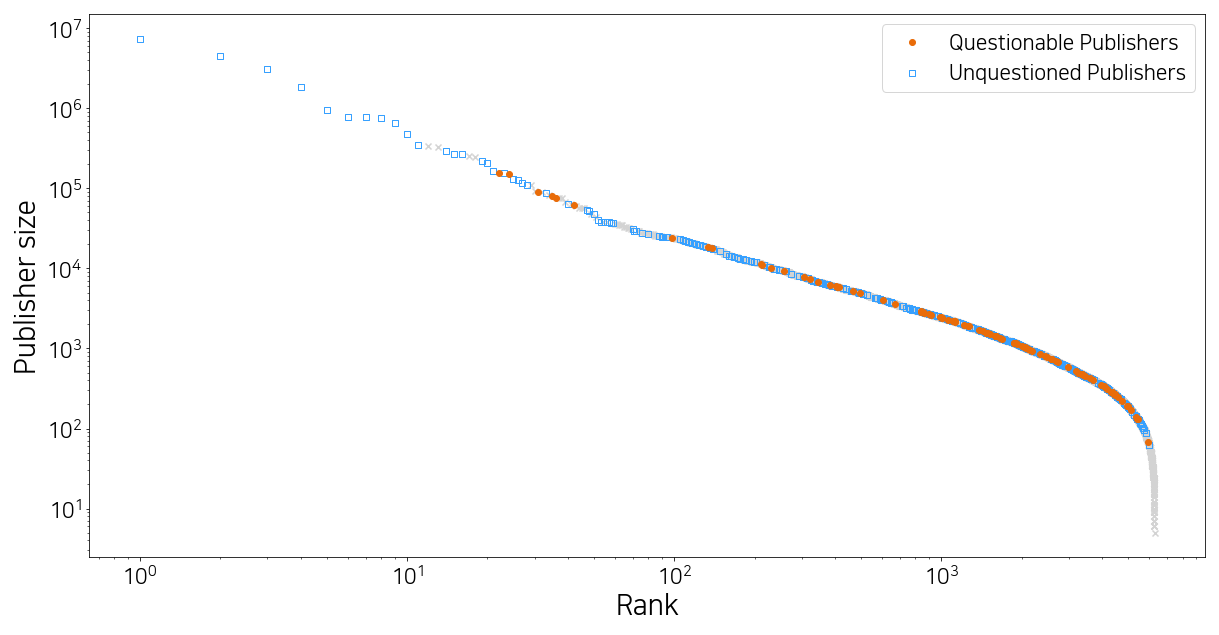}
    \caption{The rank of journals by publisher size. The publisher size is defined as the number of all publications in Scopus. We exclude articles that do not contain the publisher information. Grey dots represent publishers not belonging to any of the QJs or UJs.}
    \label{fig:sm_publisher_dist}
\end{figure}

\begin{figure}
    \centering
    \includegraphics[width=\textwidth]{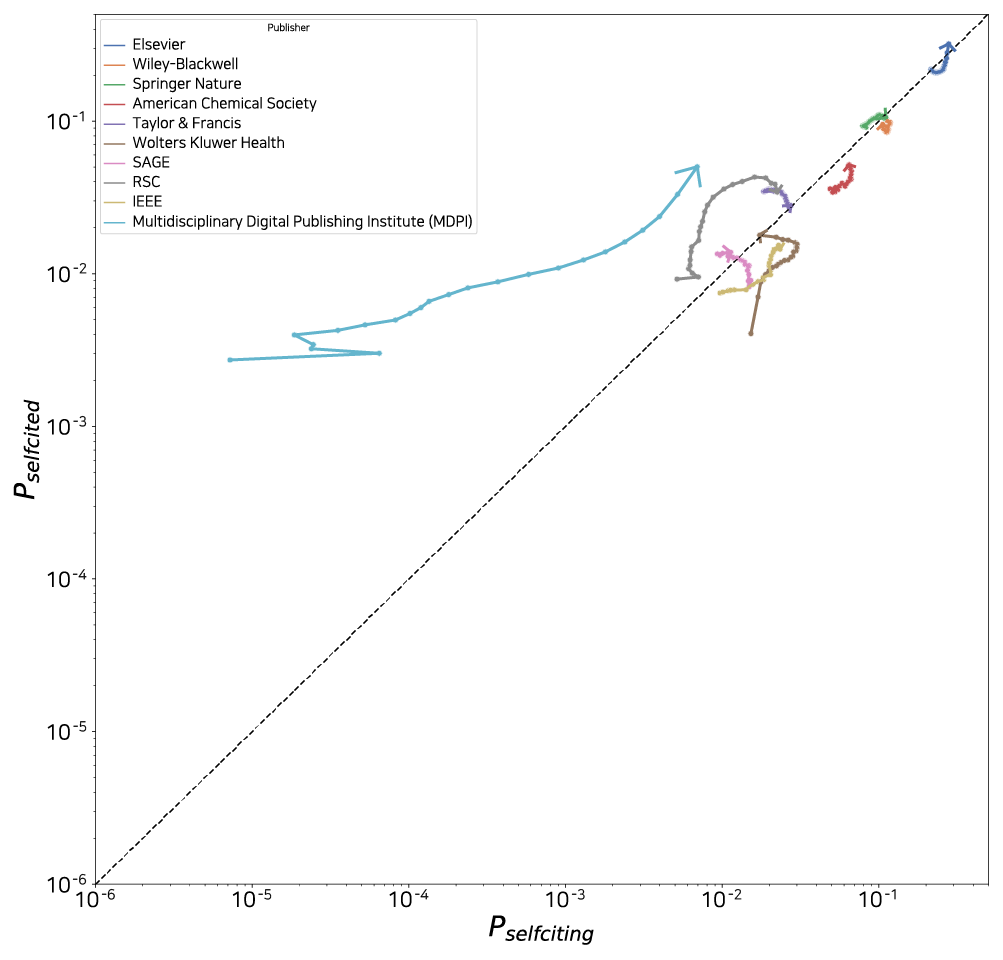}
    \caption{Fraction of publisher self-citations and self-references. The direction of the arrow indicates the temporal changes from 1996 to 2018. The largest publishers, e.g., Springer Nature, Elsevier, and Wiley--Blackwell, show more than a self-citing/self-cited rate of more than 10\%, while other publishers exhibited a lower rate.}
    \label{fig:sm_publisher_cite}
\end{figure}

\begin{figure}
    \centering
    \includegraphics[width=\textwidth]{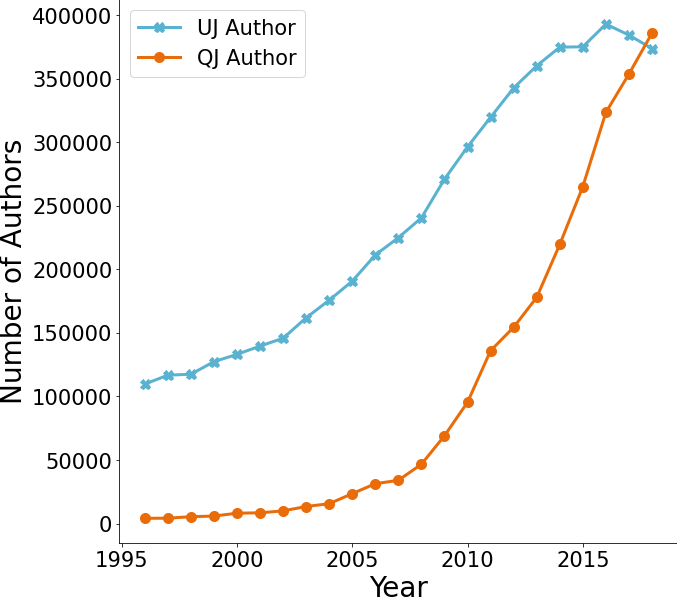}
    \caption{Number of authors who published in QJs (orange) and UJs (blue) by year.}
    \label{fig:sm_author_size}
\end{figure}

\begin{figure}
    \centering
    \includegraphics[width=\textwidth]{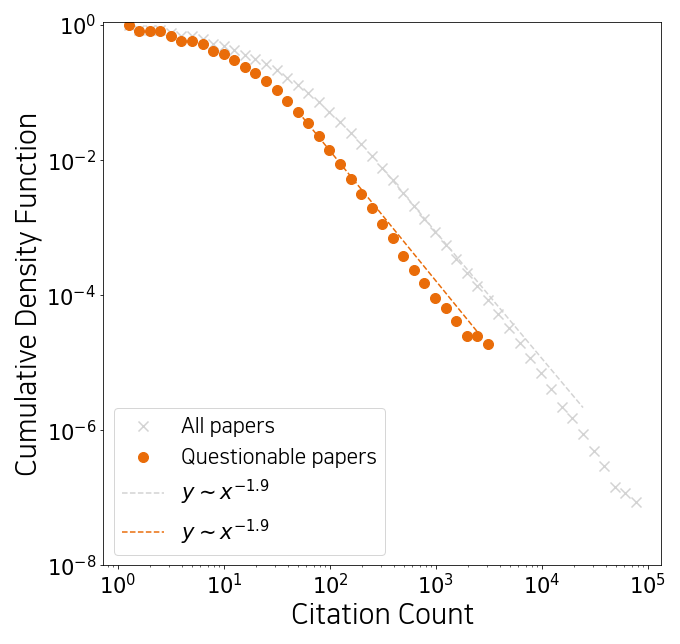}
    \caption{Distribution of publications by the number of citations. QJs have fewer publications, so they have a shorter distribution length. However, both distributions have similar tail distributions with the exponent $\alpha = 1.9$.}
    \label{fig:sm_citation}
\end{figure}

\begin{figure}
    \centering
    \includegraphics[width=0.9\textwidth]{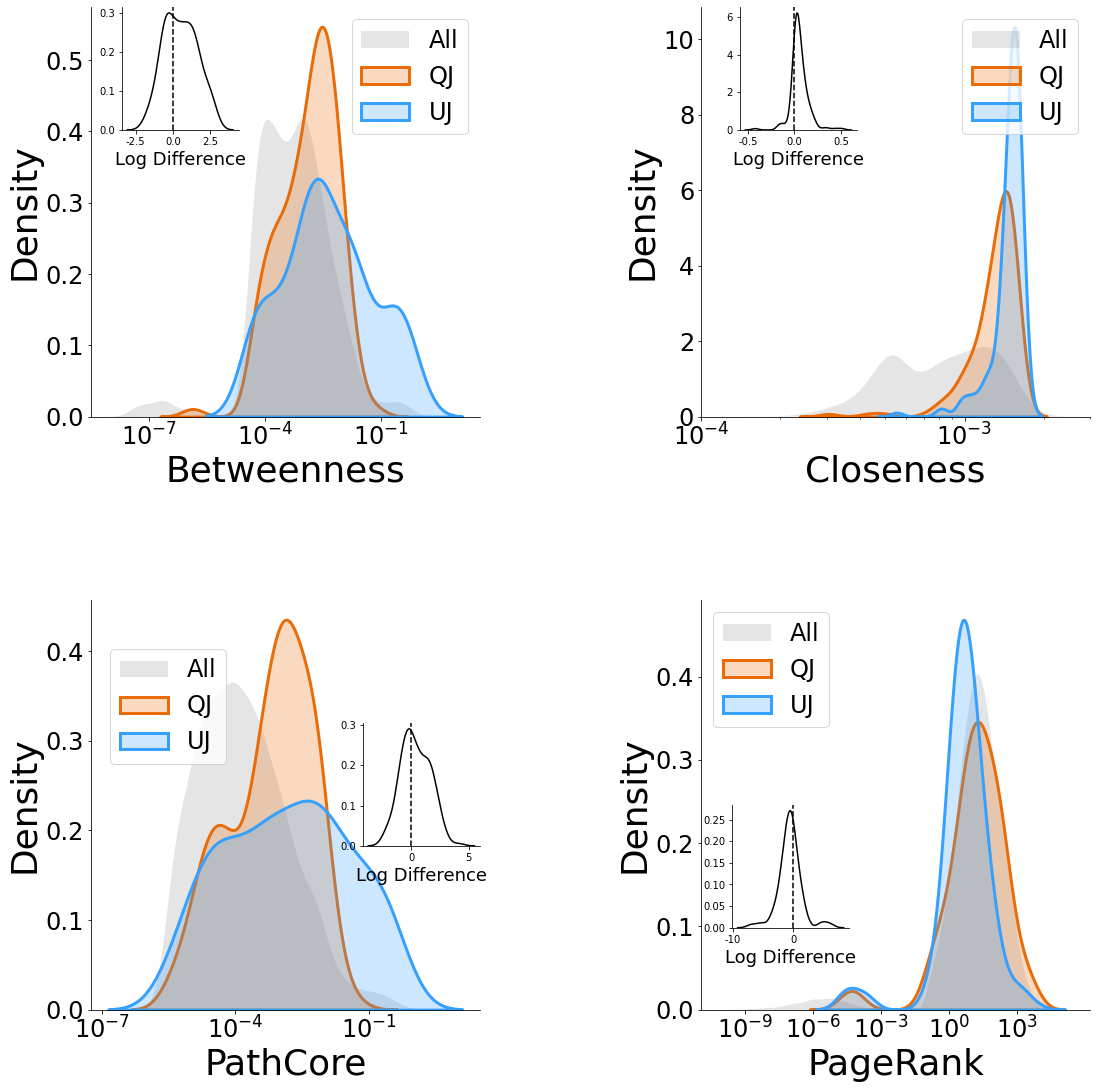}
    \caption{Centrality distribution of QJs and UJs with high journal impact ($> 3$).}
    \label{fig:sm_centrality_highjif}
\end{figure}

\begin{figure}
    \centering
    \includegraphics[width=0.9\textwidth]{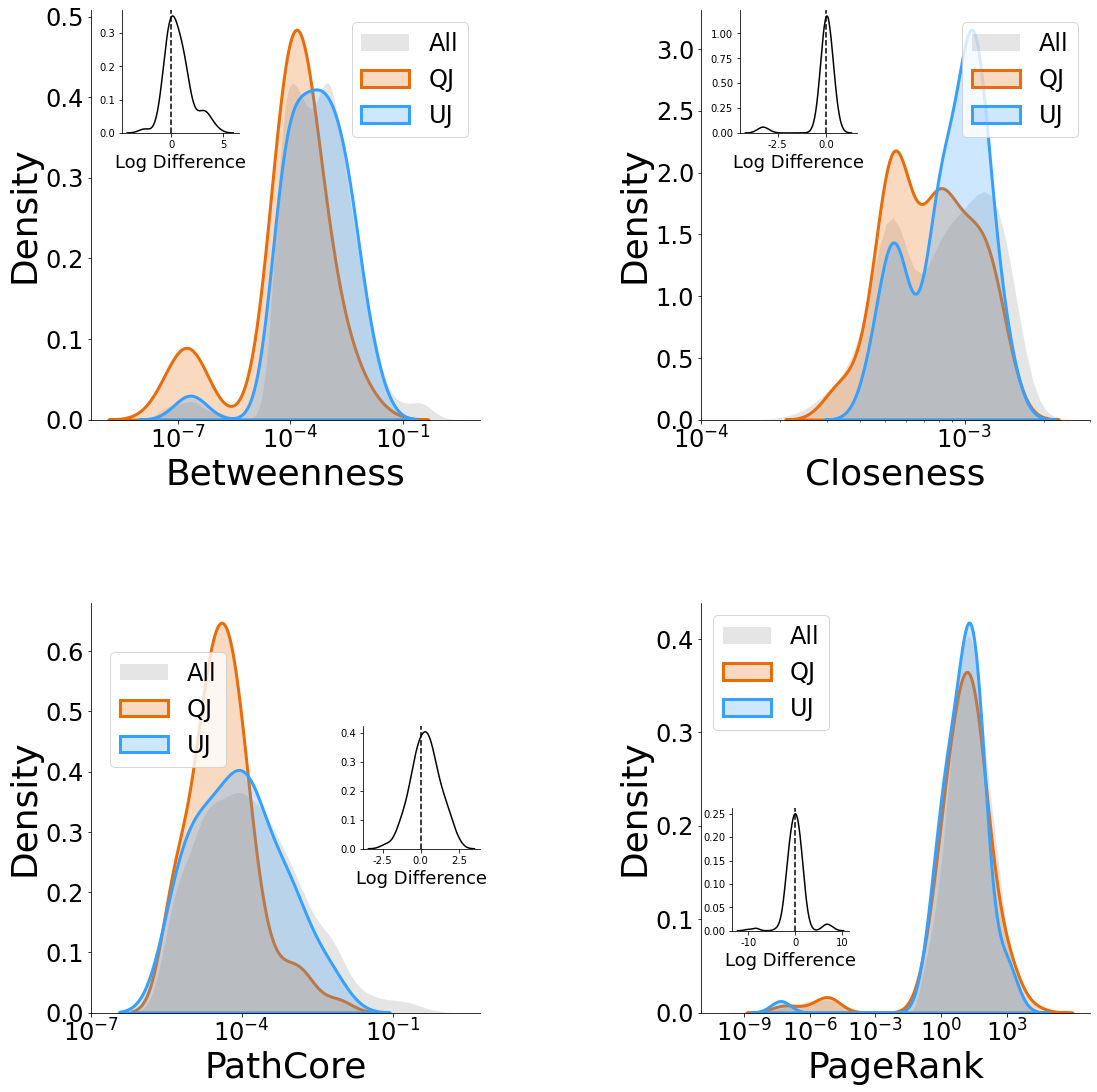}
    \caption{Centrality distribution of QJs and UJs with low journal impact ($< 0.5$).}
    \label{fig:sm_centrality_lowjif}
\end{figure}

\begin{figure}
    \centering
    \includegraphics[width=\textwidth]{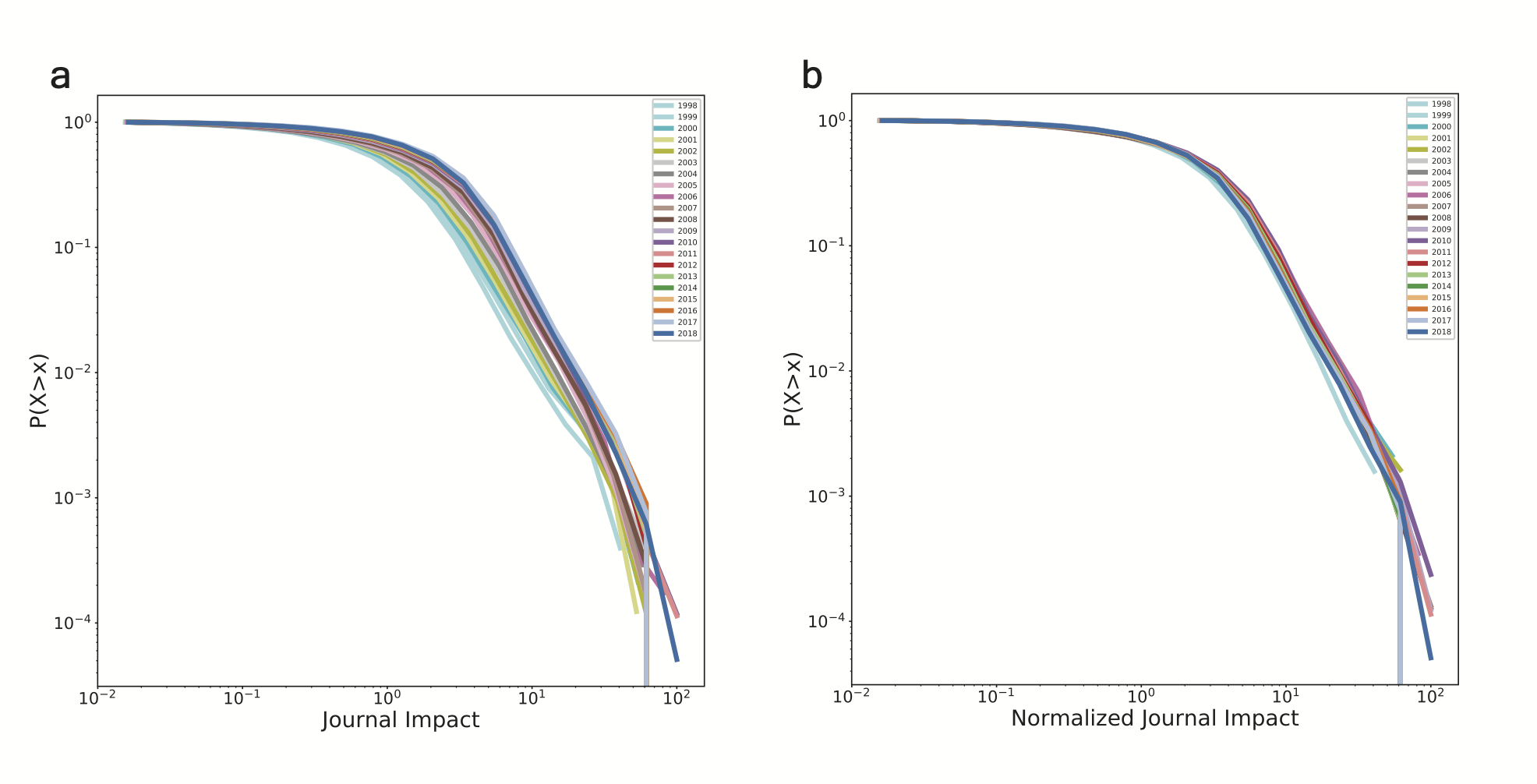}
    \caption{Result of normalising journal impact. (a) The distribution of original journal impact indicates that the impact increases annually. (b) The normalised journal impact distribution. For the normalisation, we use citations from 2017.}
    \label{fig:sm_IFnormalize}
\end{figure}

\begin{figure}
    \centering
    \includegraphics[width=\textwidth]{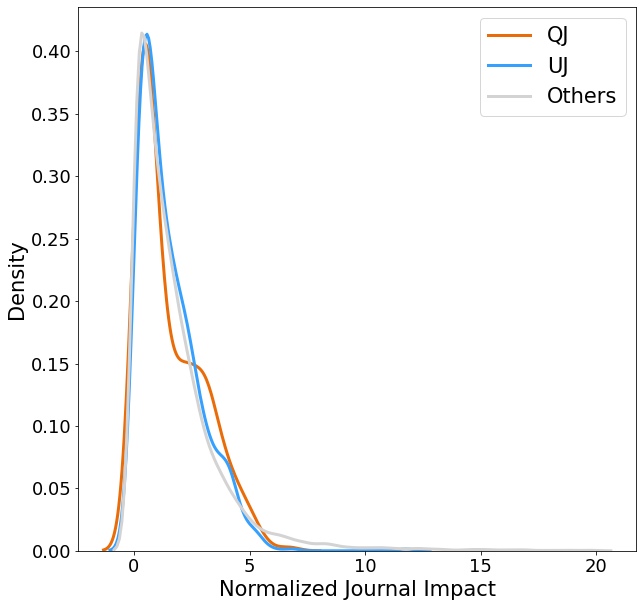}
    \caption{Distribution of normalized journal impact. Normalized journal impact is calculated for journals which have more than 30 publications in a year.}
    \label{fig:sm_impact_distribution}
\end{figure}

\noindent\textbf{This PDF file includes:}\\
\\
Supplementary text\\
Figures.~\ref{fig:sm_questionable_size} to~\ref{fig:sm_PR5y_ref}\\

\newpage
\section{Brief description of questionable journals in Beall’s list}

\begin{figure}
    \centering
    \includegraphics[width=\textwidth]{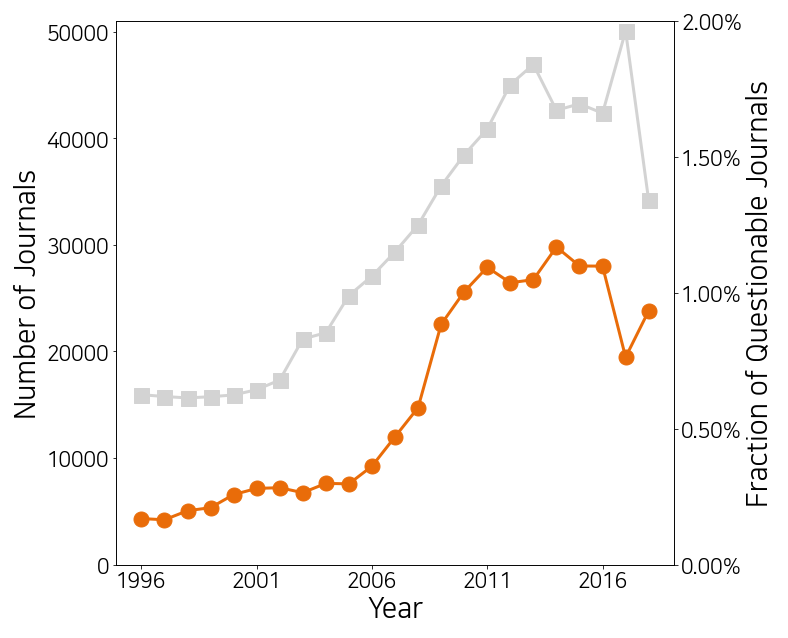}
    \caption{Number of journals (grey) published in Scopus, and the fraction of QJs in Scopus (orange). The number of journals begins to increase in 2001 and is more than 40,000 by 2011. The fraction of QJs increases in 2006 and is more than 1\% of the total journals by 2011.}
    \label{fig:sm_questionable_size}
\end{figure}

In this study, we collected 766 questionable journals (QJs), which consisted of 643 journals in the original list by Beall and 123 journals in the updated list contributed by the community. Note that QJs that publish less than 30 articles annually were excluded to prevent erroneous computations. The results revealed a massive surge in the share of QJs in 2010, as shown in Fig.~\ref{fig:sm_questionable_size}. QJs were mainly classified into medicine (31.7\%), engineering (18.1\%), biology (16.0\%), and computer science (15.6\%) based on the ASJC of Scopus. Some QJs were simultaneously classified into multiple subject categories. For instance, journals belonging to biology were occasionally co-classified under medicine. Only $25\%$ of the QJs published 100 or more articles annually; therefore, correlation between journal impact and publication volume was insignificant. The correlation between annual publication size and journal impact was only 0.22.

\section{Note on selection criteria of unquestioned journals}

\subsection{Publication size}
The ecosystem of mega journals is different from that of small journals reflected in the publication size. We varied the annual publication size of journals for the selection criteria to compensate for the impact of journal size. However, it is impossible to match the journal impact. On the one hand, if the margins for matching the journal size were set too narrow, no journal would have fit the journal impact. On the other hand, if the margins were too broad, the UJs might have a different publication policy regarding the corresponding QJs.

\begin{figure}
    \centering
    \includegraphics[width=0.7\textwidth]{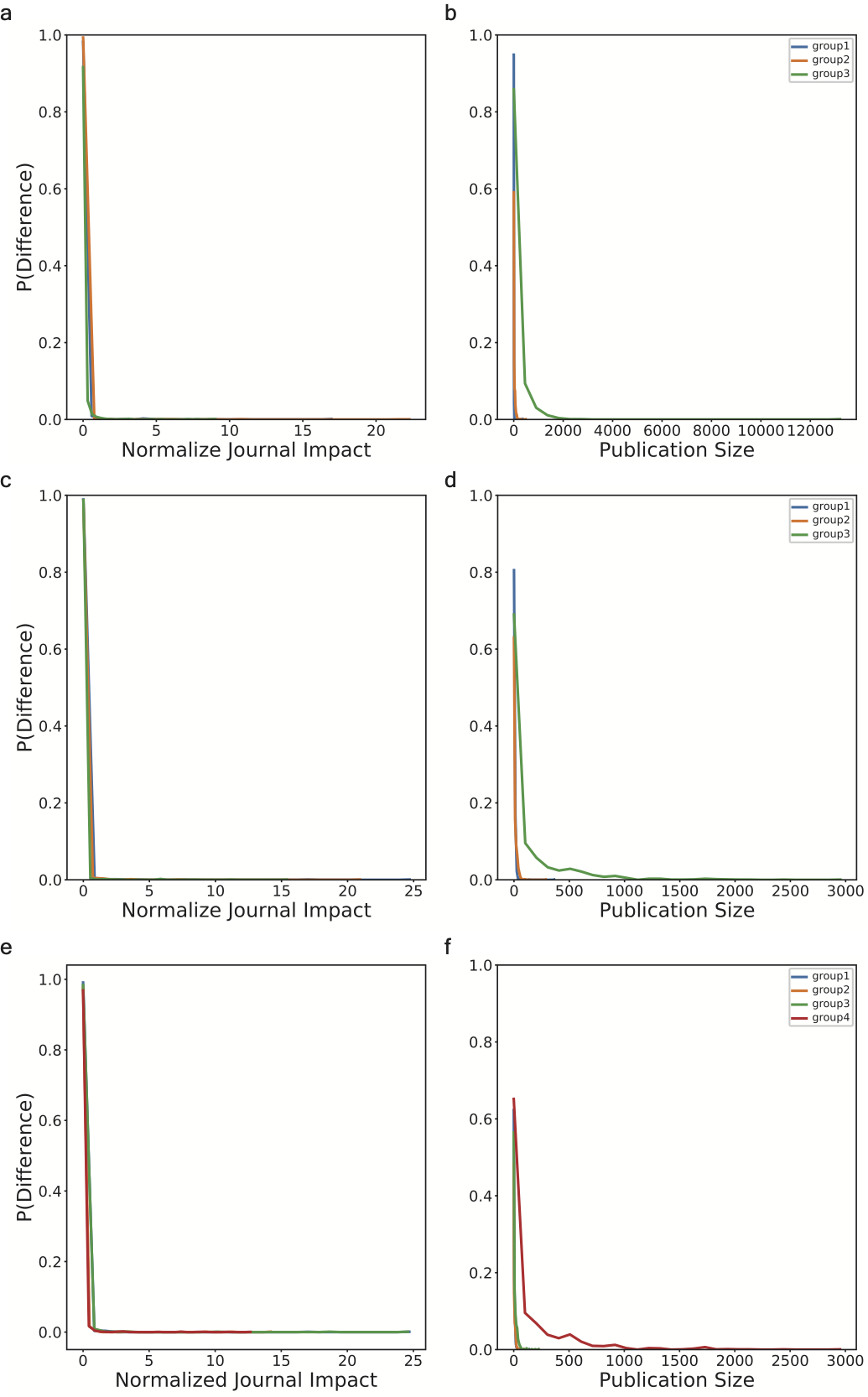}
    \caption{Difference in journal impact and publication size between QJs and UJs for different size criteria: (a) and (b) $\pm \sigma$ with a log scale, (c) and (d) 33rd percentiles, and (e) and (f) quartiles. The difference in publication size is the highest for $\pm \sigma$ with a log scale, and the difference in the normalised journal impact is higher for the quartile than for the 33rd percentile.}
    \label{fig:sm_controlgroup}
\end{figure}

In this study, we tested three possible candidate metrics to calculate the adequate division of journal size that will minimise the differences in journal size and impact: (a) $\pm \sigma$ with a log scale, (b) 33rd percentiles, and (c) quartiles (Fig.~\ref{fig:sm_controlgroup}). We chose the candidates of UJs based on the selection criteria with the three types of journal size matching. For a subject category, the publication size followed a heavy-tailed distribution. Thus, we considered the division of $\pm \sigma$ with a log scale as the first candidate. However, using (a) $\pm \sigma$ with a log scale resulted in a larger annual publication size difference than that obtained using the other metrics (Fig.~\ref{fig:sm_controlgroup}). Both (b) 33rd percentiles and (c) quartiles yielded better results than (a). In contrast, the division of 33rd percentiles had smaller differences in journal impact and size. Therefore, we concluded that 33rd percentiles can calculate the adequate division. 

\subsection{Additional analysis using metrics of the Journal Citation Reports}
\begin{figure}
    \centering
    \includegraphics[width=0.8\textwidth]{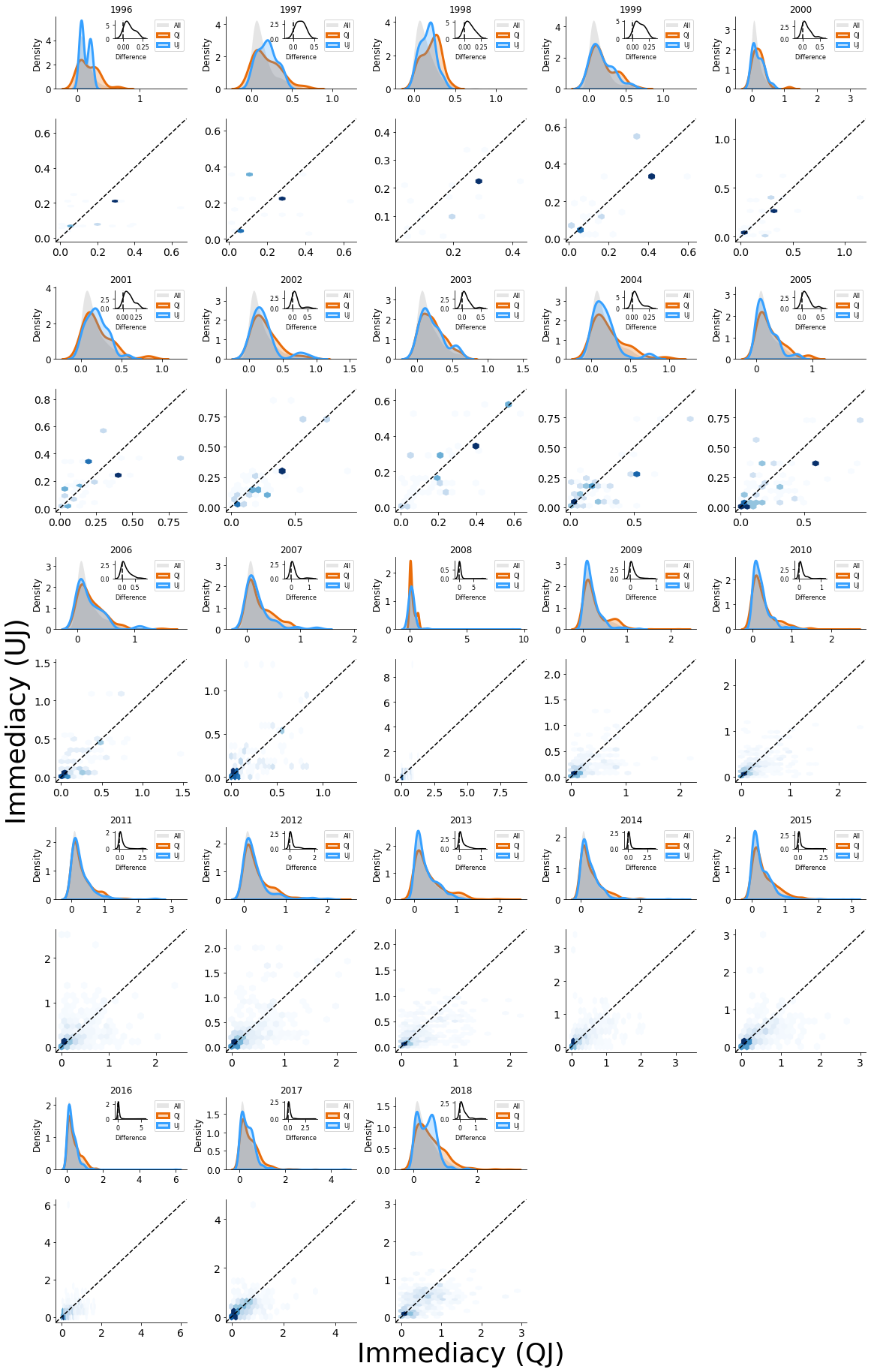}
    \caption{Distributions and interrelations of immediacy index between QJs (horizontal axis) and UJs (vertical axis). The dashed black line represents the linear relationship of $y=x$. The distribution above the heatmap is the kernel density estimation of the immediacy index for all journals, QJs, and UJs, whereas its inset displays the difference between the immediacy indices of UJs and QJs.}
    \label{fig:sm_immediacy}
\end{figure}

\begin{figure}
    \centering
    \includegraphics[width=0.8\textwidth]{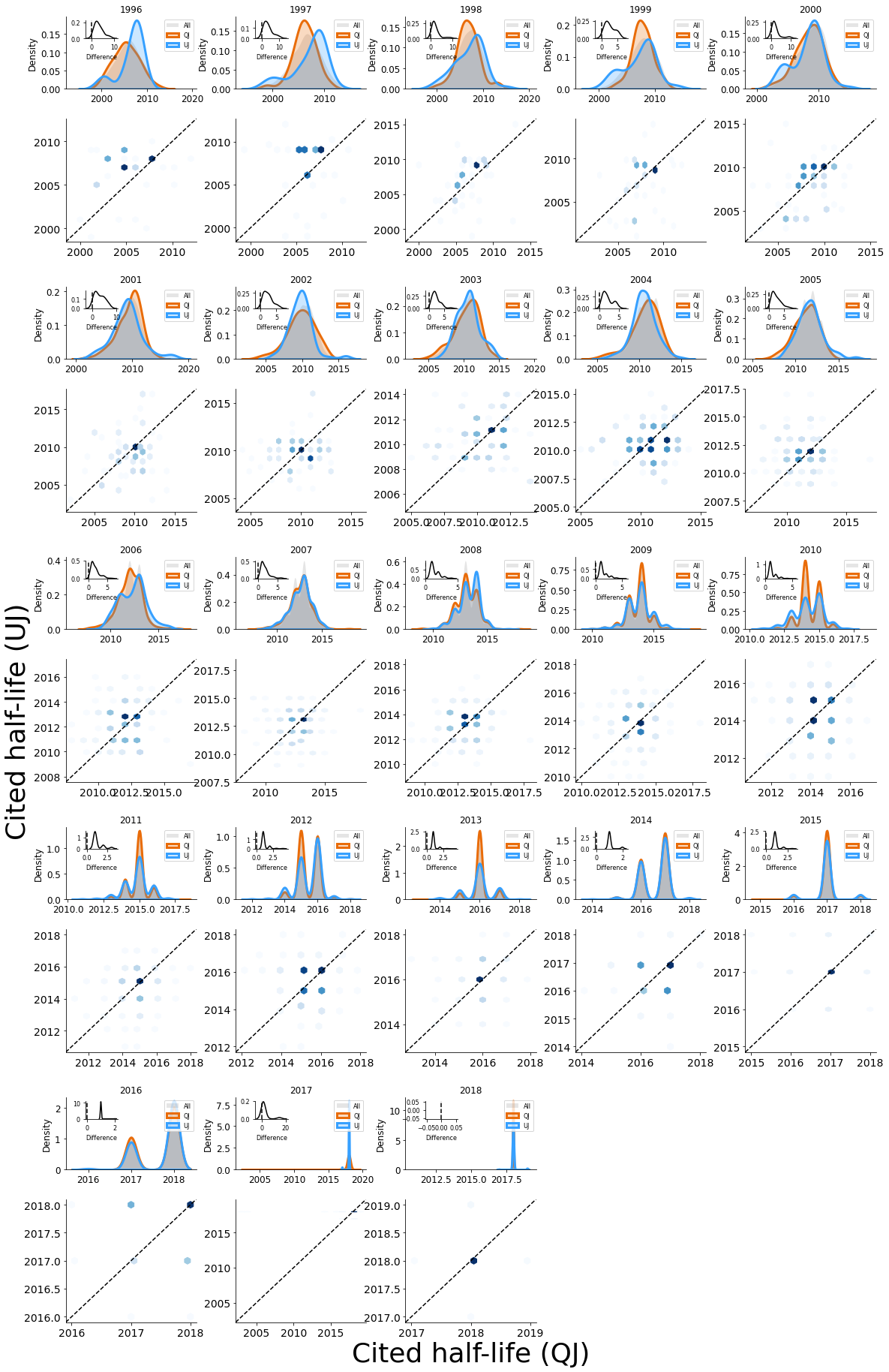}
    \caption{Distributions and interrelations of cited half-life between QJs (horizontal axis) and UJs (vertical axis). The dashed black line represents the linear relationship of $y=x$. The distribution above the heatmap is the kernel density estimation of the cited half-life for all journals, QJs, and UJs, whereas its inset displays the difference between the cited half-lives of UJs and QJs.}
    \label{fig:sm_halflife_cited}
\end{figure}

\begin{figure}
    \centering
    \includegraphics[width=0.85\textwidth]{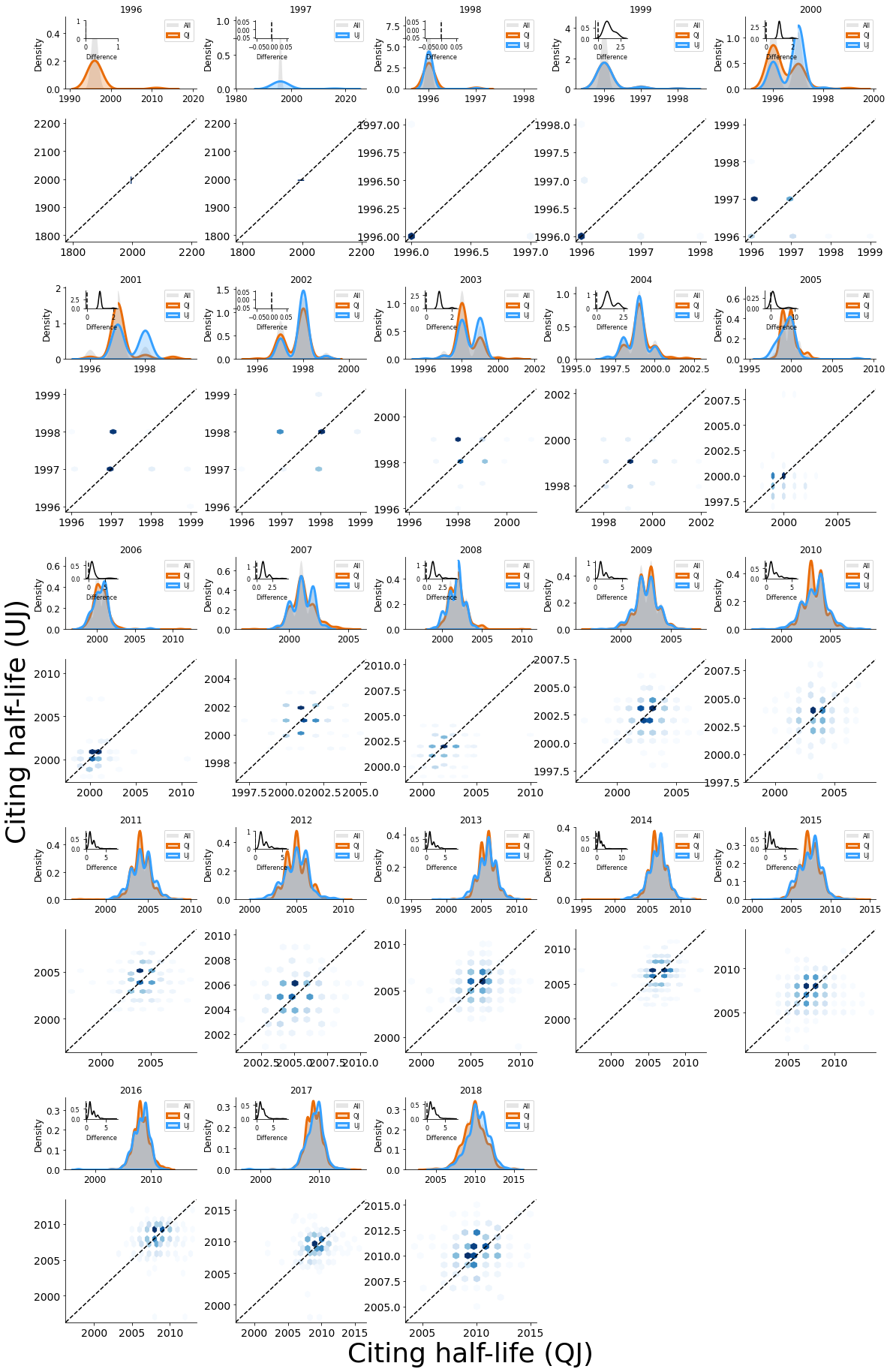}
    \caption{Distributions and interrelations of citing half-life between QJs (horizontal axis) and UJs (vertical axis). The dashed black line represents the linear relationship of $y=x$. The distribution above the heatmap is the kernel density estimation of the citing half-life for all journals, QJs, and UJs, whereas its inset displays the difference between the cited half-lives of UJs and QJs.}
    \label{fig:sm_halflife_citing}
\end{figure}

We also tested three well-known metrics used by the JCR to compare the citation characteristics of QJs and select UJs. The immediacy index was used to measure the speed of a citation by computing the average citation count in the year of publication (Fig.~\ref{fig:sm_immediacy}). The cited half-life was used to estimate the lifetime of an article, which was measured by the median age of the forward citations (Fig.~\ref{fig:sm_halflife_cited}). The citation half-life was used to estimate the age of the topic for an article by the median age of the backward citations (Fig.~\ref{fig:sm_halflife_citing}). These results showed that most journals, QJs and UJs, had similar indices. To conclude, UJs do not significantly differ from QJs in terms of the citation timescale.

\section{Additional comparisons with other journals in Scopus}
\begin{figure}
    \centering
    \includegraphics[width=\textwidth]{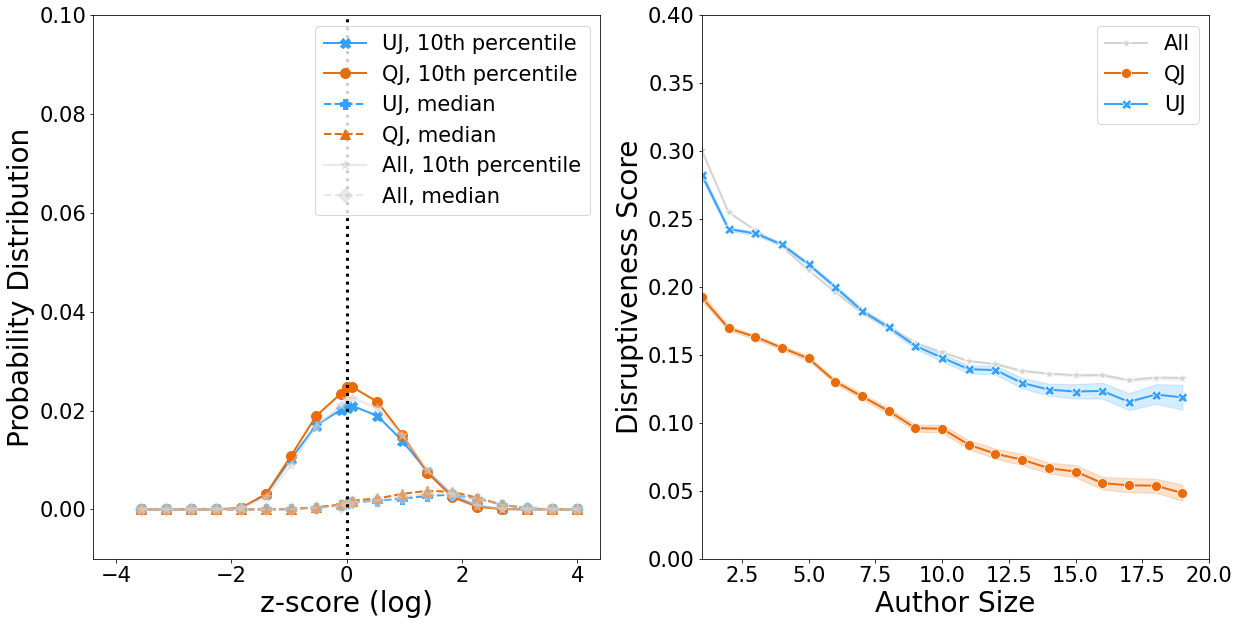}
    \caption{Distribution of novelty and disruptiveness for each publication. (a) Distribution of novelty by reference combinations. We plot the 10th percentile and median as the representative z-score. (b) The average disruptiveness score is displayed by the number of authors.}
    \label{fig:sm_paper_measures}
\end{figure}

\begin{figure}
    \centering
    \includegraphics[width=\textwidth]{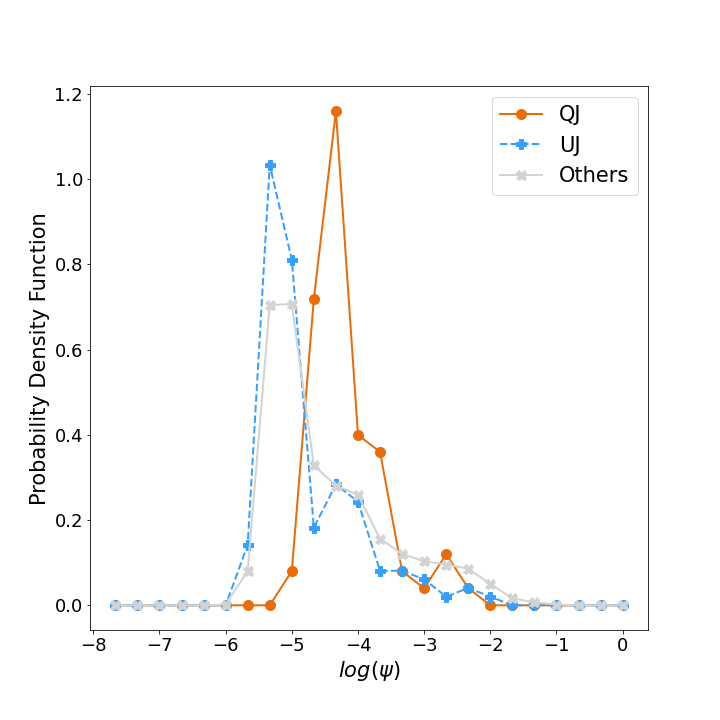}
    \caption{Distribution of $\psi$. The $\psi$ is calculated for 9,906 journals that have publisher information. The distribution shows that UJs and other journals show similar $\psi$ for the modes, while the mode of QJs is located at a higher $\psi$.}
    \label{fig:sm_psi_distribution}
\end{figure}

We compared QJs, UJs, and all other journals to test the validity of our unquestioned journals. First, the disruptiveness and novelty of all other publications were estimated. The other journals and unquestioned journals show similar distributions for disruptiveness index and novelty; however, questionable publications show different distributions (Fig.~\ref{fig:sm_psi_distribution}). We then compare the $\psi$ for the three sets of journals. The publisher information is required to calculate the $\psi$, which consists of more than two journals, thus, $\psi$ can only be estimated for 9,906 journals. The distribution demonstrates that UJs and other journals have a closely located mode, whereas the mode of QJs is located significantly  above than that of the other sets (Fig.~\ref{fig:sm_psi_distribution}).Taken together with the results from Fig.~\ref{fig:journal}, Fig.~\ref{fig:sm_paper_measures} and Fig.~\ref{fig:sm_psi_distribution}, we can summarise that QJs show different patterns than UJs and other journals, while UJs and other journals distribute similar patterns. Thus, although UJs are rather a small subset of the entire Scopus, they can be representative of all other journals in the Scopus.

\section{Validation of publication solidarity index}
\label{sec:sm_psi_validation}
We proposed a publication solidarity index ($\psi$) in eq.~\ref{eq:psi} to compute self-favouring journal citations for the publishers. The $\psi$ encompasses a journal's self-favourable citations for its publisher, the publisher's average self-citations, and the number of publications. The measure rises when a journal gives more citations to its publisher since the behaviour is more aggressive than the other publishers. The number of publications is considered as a null model under the hypothesis that one randomly cites papers with equal possibility. We, thus, use this denominator to reduce the impact of the publishers’ size.
 
In this section, we performed three extra analyses, using synthetic citation counts, to confirm the validity of the $\psi$. First, we explored a scenario in which a journal provides its publisher with a large number of citations. When the number of journal citations ($\sum R_r$) changes, the number of self-citations within the publisher ($Q_r$ and $Q_c$) changes accordingly, and thus, $\sum R_r = Q_r = Q_c$ (see Supplementary Fig.~\ref{fig:sm_psi_validation}(a)). Second, a journal generated citations for the publisher, although this only occupied a minor portion of the publisher's self-citations. The quantity of the journal citations varied, but self-citation rates ($Q_r$ and $Q_c$) remained constant (see Supplementary Fig.~\ref{fig:sm_psi_validation}(b)). Third, a journal obtained citations solely from other journals from its publisher, therefore, only $\sum R_c$ varies (see Supplementary Fig.~\ref{fig:sm_psi_validation}(c)). The first two scenarios depict a journal providing the publisher with an increasing number of self-citations, whereas the third case depicts the publisher giving citations to the journal. The results show that the $\psi$ increases when the number of self-citations to its publisher increases. We also find that the peak height is smaller when the publisher's self-citing behaviour stays constant and insignificant for the total publisher's self-citation levels (compare Supplementary Fig.~\ref{fig:sm_psi_validation}(a) and (b)). This means that if a journal cites its publisher constantly and frequently, and especially it is a considerable portion of the publisher's self-citations, $\psi$ becomes higher. When the number of citations received from the publisher increases, on the other hand, $\psi$ decreases. This means that the focus should be on capturing a journal's aggressive citing activity rather than on the number of citations it receives. In conclusion, the metric successfully captures a journal's self-favouring citations, whereas external influences have little effect on the trend.

\begin{figure}
    \centering
    \includegraphics[width=\textwidth]{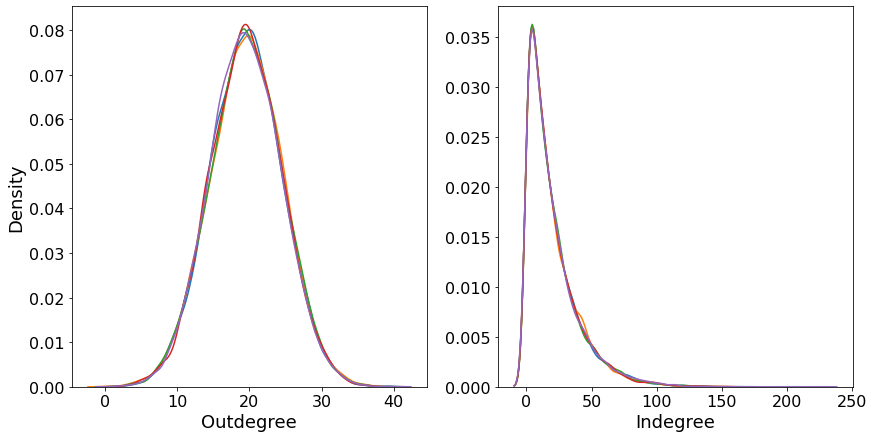}
    \caption{Degree distribution of generated network. Gaussian out-degree distribution $N(20, 5)$ and power-law in-degree distribution with an exponent of 3 using a preferential attachment.}
    \label{fig:sm_psi_degree}
\end{figure}

To provide a better interpretation of $\psi$, we employ a synthetic network approach. We construct a random citation network of papers with a configuration model~\cite{molloy1995critical} and compute $\psi$ with citation rewiring. First, we set 5 components as publishers, where each component has 5 subsets as journals. The size of components is randomly drawn from a uniform distribution between 450 and 550. We assign one of five journals for each node based on its belonging component, and thus each journal is composed of 100 nodes on average. We then randomly connect the nodes with gaussian out-degree distribution $N(20, 5)$ and power-law in-degree distribution with an exponent of 3 using a preferential attachment (Fig.~\ref{fig:sm_psi_degree}). The generated citation distributions are intended to emulate real-world citation distribution. All five randomly generated components have a similar number of citations, self-citations, journal impact, and $\psi$ at first.

We then rewire the links using the following rules. First, we choose a link for rewiring at random. We reassign the target node (as cited by the link) to a different node while keeping the source node to maintain the out-degree for a node. The chance of a node being selected as a new target node is determined by its journal. We assign a journal with a special probability of publisher-level self-citation for each component, randomly chosen among $1/2$, $1/4$, $1/8$, $1/16$, and $1/16$. All other journals should have a probability of $1/5$, as they have no publisher preference. A journal with a publisher self-citation rate of $r$ may choose a new paper in the same publisher to cite with a probability of $r$. As a result, a greater probability indicates that a node will cite more papers from the same publisher during the rewiring process. We generate twenty random networks and calculate the ensemble-averaged $\psi$ for journals with a special probability.

\begin{figure}
    \centering
    \includegraphics[width=\textwidth]{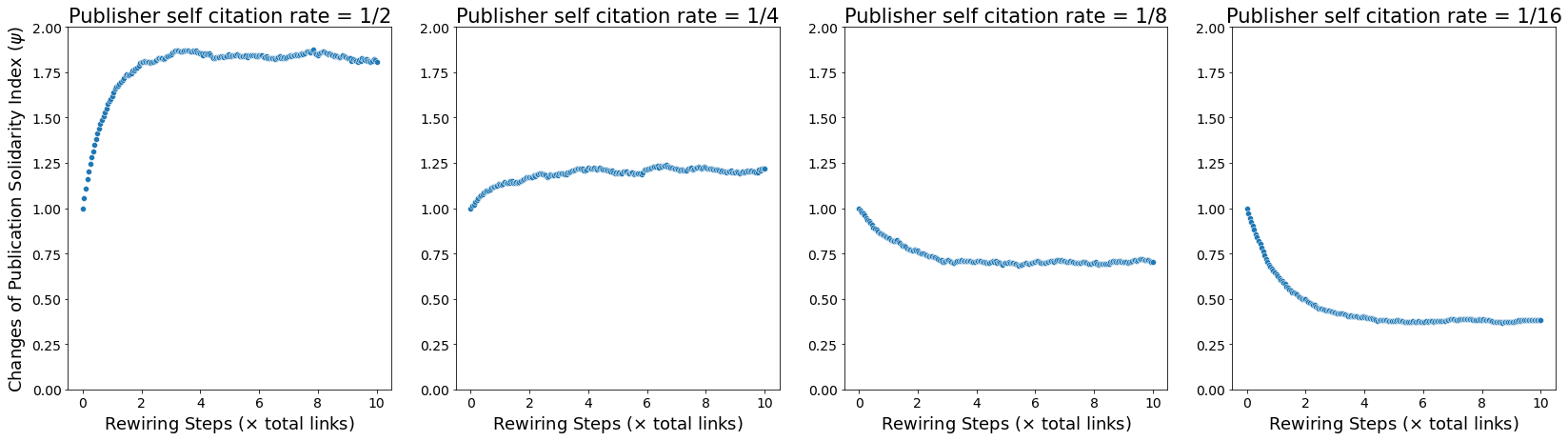}
    \caption{The change of the publication solidarity index ($\psi$) after the citation rewiring process. The ratio is calculated by dividing the $\psi$ of rewired networks by those of the original network and averaging over 20 ensembles. The publisher self-citation rates of target journals are set between $1/2$ and $1/16$ during each rewiring process, whereas the self-citation rates of other journals are fixed at $1/5$.}
    \label{fig:sm_psi_change}
\end{figure}

The analysis demonstrates clearly that $\psi$ increases when the publisher's self-citation rate is high (Fig.~\ref{fig:sm_psi_change}). We observed an increment of $\psi$ for two journals whose publishers self-cite at a higher rate than the average. Meanwhile, journals with a lower rate of publisher self-citations demonstrate a decrease of $\psi$, as they rarely cite papers published by the same publisher. The shift is greater in the early stages, implying that a small number of changes in publisher self-citations would result in a great increment of $\psi$. Thus, our measure can capture the anomaly for the early stage of malpractice using publisher-level self-citations. In addition, when we rewire more than twice of total links, the $\psi$ is saturated and is almost constant.

\section{Additional analysis with disruptiveness index}

\begin{figure}
    \centering
    \includegraphics[width=\textwidth]{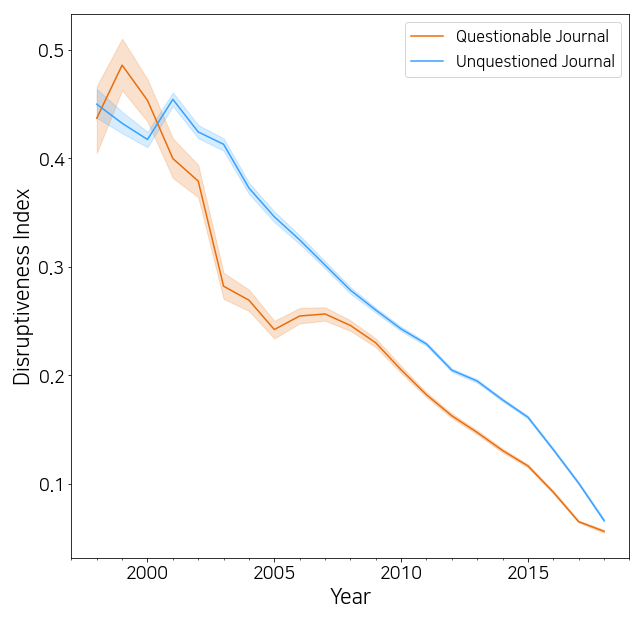}
    \caption{Average disruptiveness index by year. The publications in QJs have a higher disruptiveness index than those in UJs, except from 1998 to 2000.}
    \label{fig:sm_disruptiveness_paper_year}
\end{figure}

We conducted further comparisons between QJs and UJs in terms of the disruptiveness index. First, we computed the average disruptiveness index for a year (Fig.~\ref{fig:sm_disruptiveness_paper_year}). The UJs consistently scored a higher disruptiveness index, except during 1998--2000. There were only a few QJs in the late 1990s; thus, the index may fluctuate strongly.

\begin{figure}
    \centering
    \includegraphics[width=\textwidth]{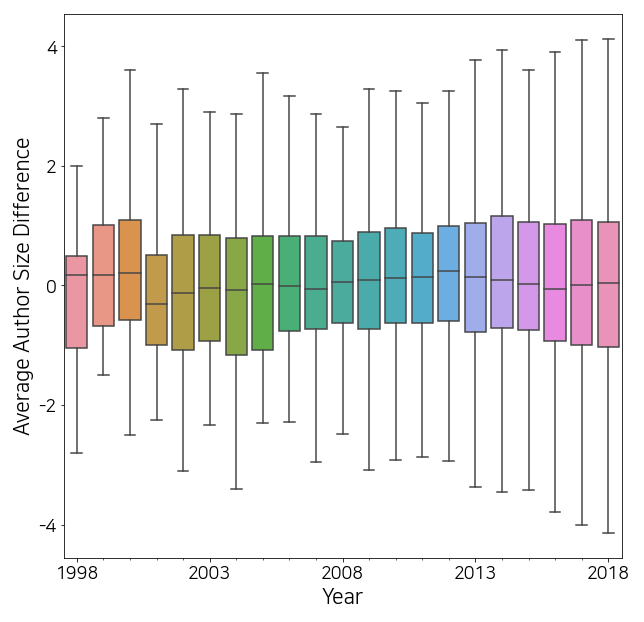}
    \caption{Difference in the average author size between QJs and UJs. We calculate the difference by subtracting the mean number of authors for the QJs from that of authors for the UJs. The distribution by year demonstrates the nearly identical differences between the journals.}
    \label{fig:sm_disruptiveness_authorsize}
\end{figure}

\begin{figure}
    \centering
    \includegraphics[width=\textwidth]{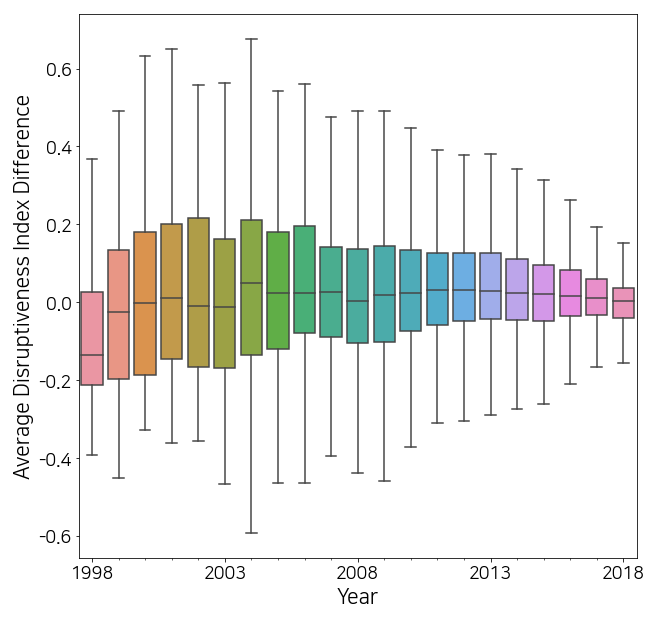}
    \caption{Difference in the average disruptiveness index between the QJs and UJs. We calculate the difference by subtracting the disruptiveness index of publications in QJs from those of UJs. Thus, the difference is high when a UJ shows a higher disruptiveness index than a QJ.}
    \label{fig:sm_disruptiveness_difference}
\end{figure}

\begin{figure}
    \centering
    \includegraphics[width=\textwidth]{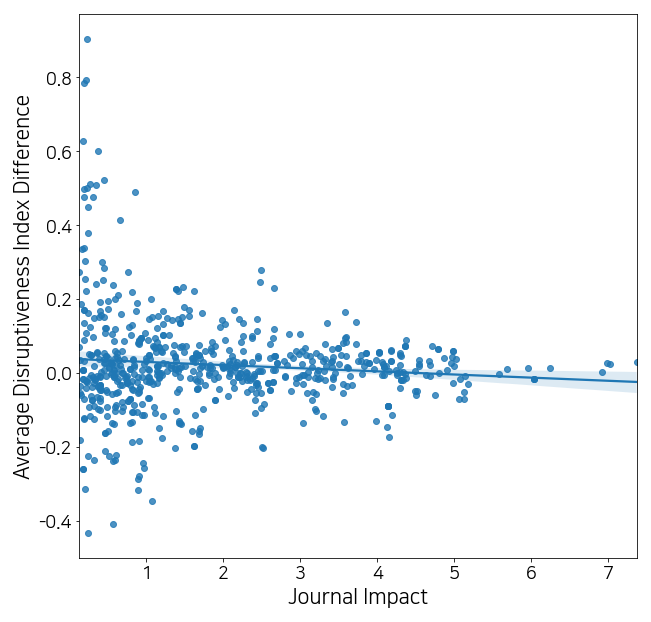}
    \caption{Difference in average disruptiveness index between QJs and UJs by journal impact. The scatterplot shows that high-impact journals fluctuate less, whereas low-impact journals fluctuate more. The linear regression shows that journal impact does not influence the difference in the disruptiveness index.}
    \label{fig:sm_disruptiveness_IF}
\end{figure}

We also found no difference in the disruptiveness index at the journal level. The differences in the average number of authors (Fig.~\ref{fig:sm_disruptiveness_authorsize}), average disruptiveness index (Fig.~\ref{fig:sm_disruptiveness_difference}), and journal impact (Fig.~\ref{fig:sm_disruptiveness_IF}) revealed an insignificant margin between QJs and UJs. Specifically, the QJs and UJs had similar number of authors and disruptiveness index on average. The average disruptiveness index fluctuated more for journals with smaller impacts, while other journals showed only a small difference. To summarise, while the publication-level disruptiveness index is high for UJs, the journal-level disruptiveness indices are the same for both QJs and UJs.

\section{Robustness of constructed network}

\begin{figure}
    \centering
    \includegraphics[width=0.85\textwidth]{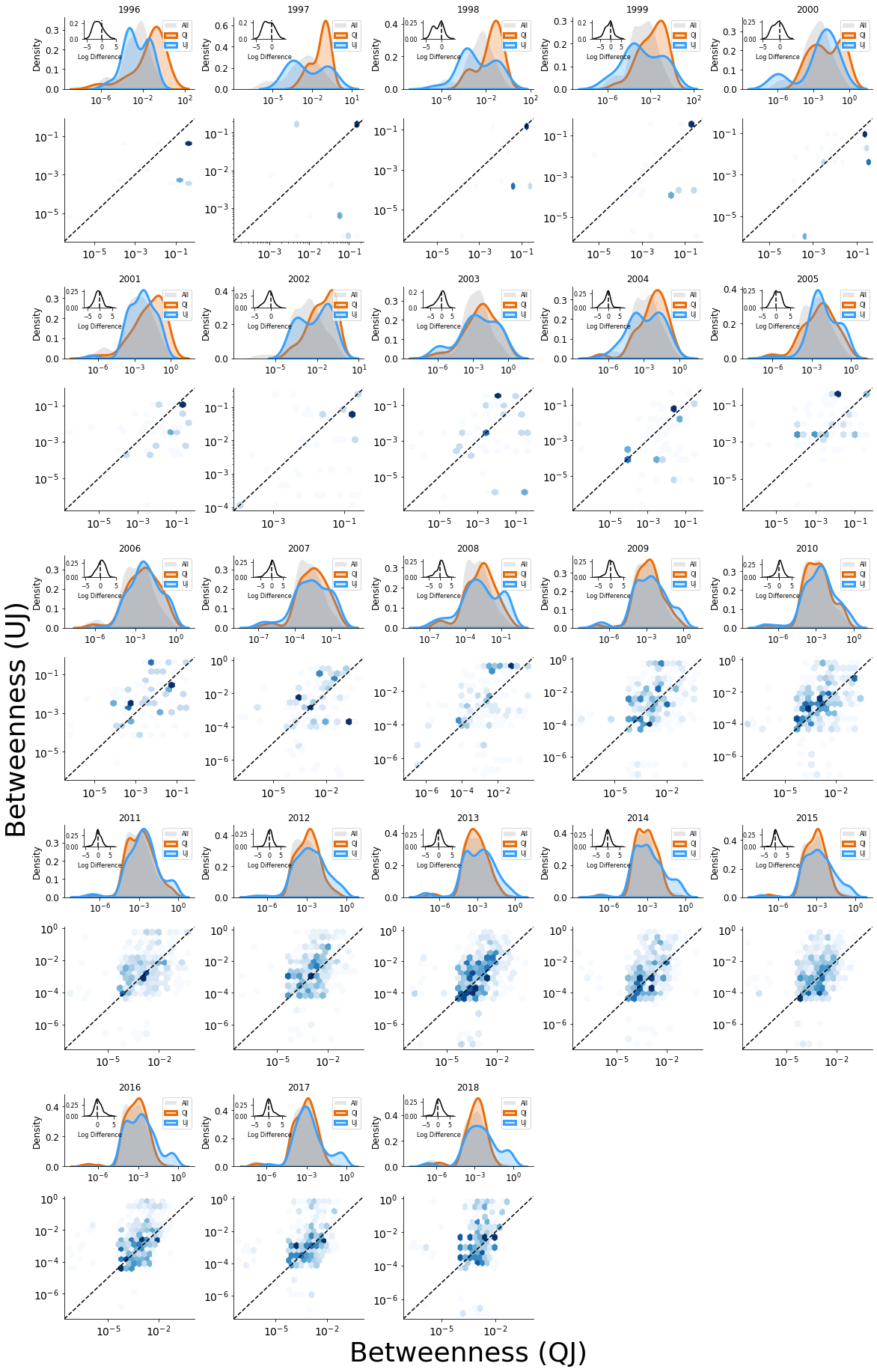}
    \caption{Betweenness centrality computed in the journal citation network with two years of citation time window.}
    \label{fig:sm_BC2y}
\end{figure}

\begin{figure}
    \centering
    \includegraphics[width=0.85\textwidth]{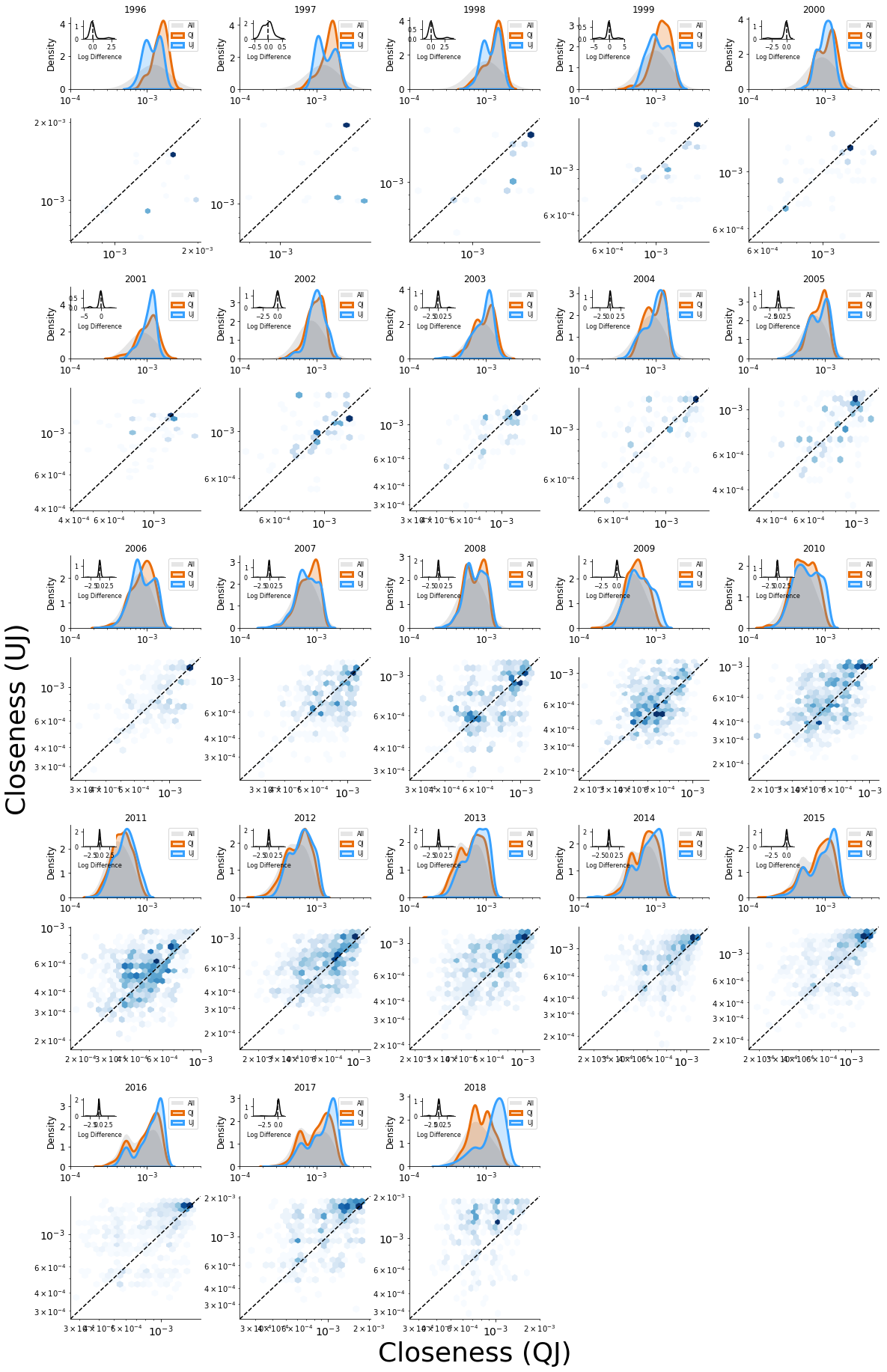}
    \caption{Closeness centrality computed in the journal citation network with two years of citation time window.}
    \label{fig:sm_CC2y}
\end{figure}

\begin{figure}
    \centering
    \includegraphics[width=0.85\textwidth]{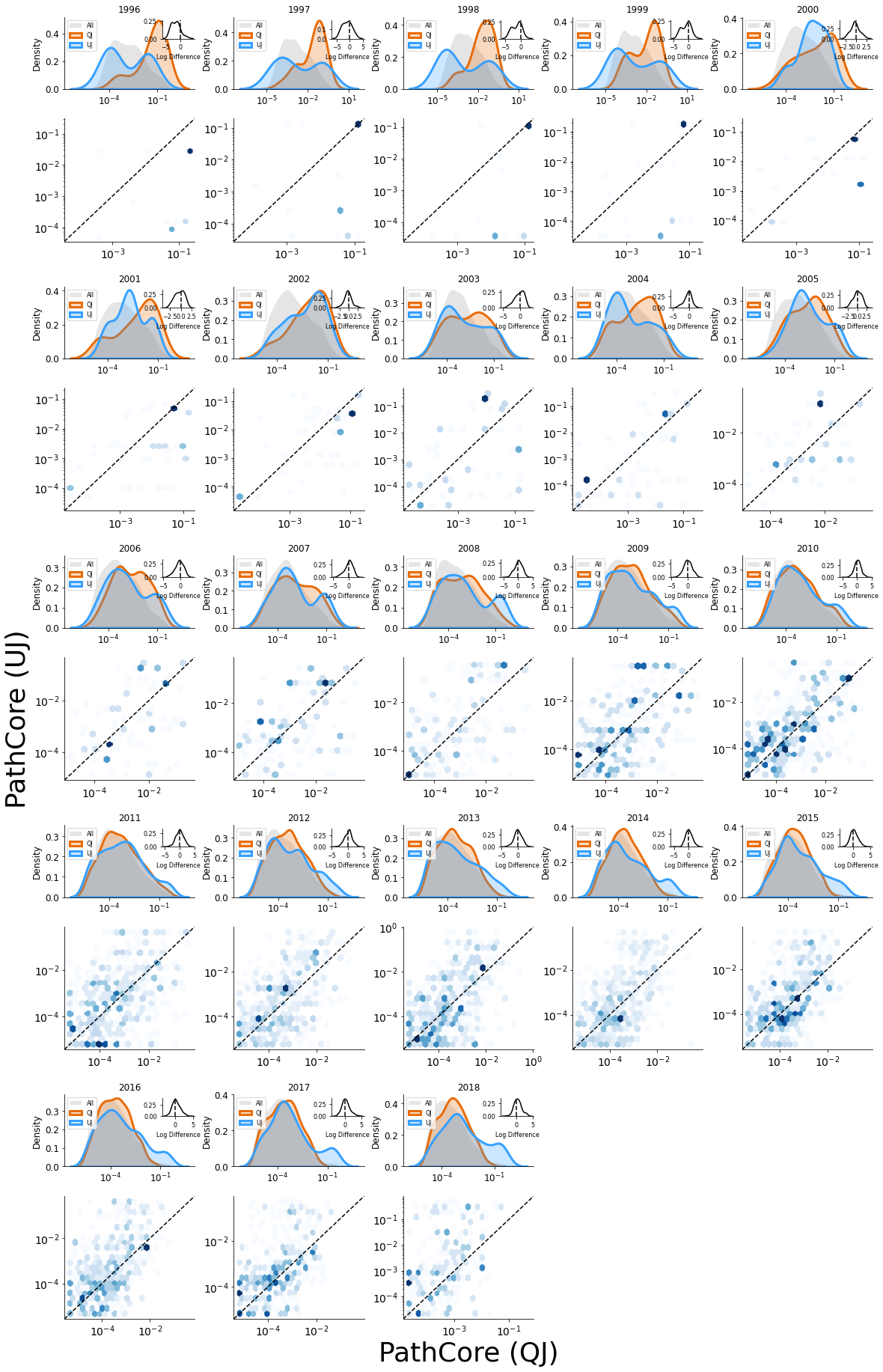}
    \caption{PathCore score computed in the journal citation network with two years of citation time window.}
    \label{fig:sm_PC2y}
\end{figure}

\begin{figure}
    \centering
    \includegraphics[width=0.85\textwidth]{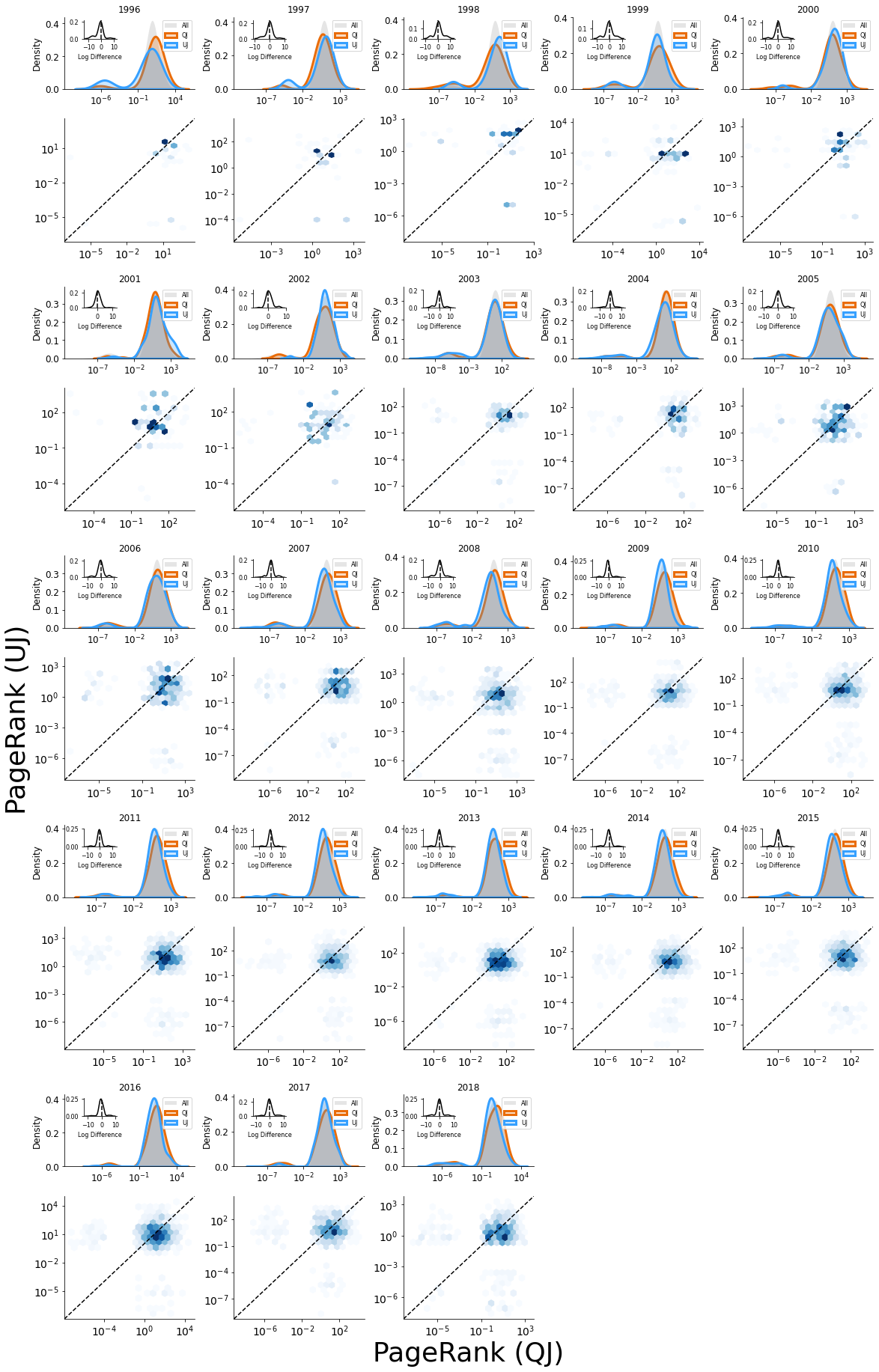}
    \caption{PageRank centrality computed in the journal citation network with two years of citation time window.}
    \label{fig:sm_PR2y}
\end{figure}

\begin{figure}
    \centering
    \includegraphics[width=0.85\textwidth]{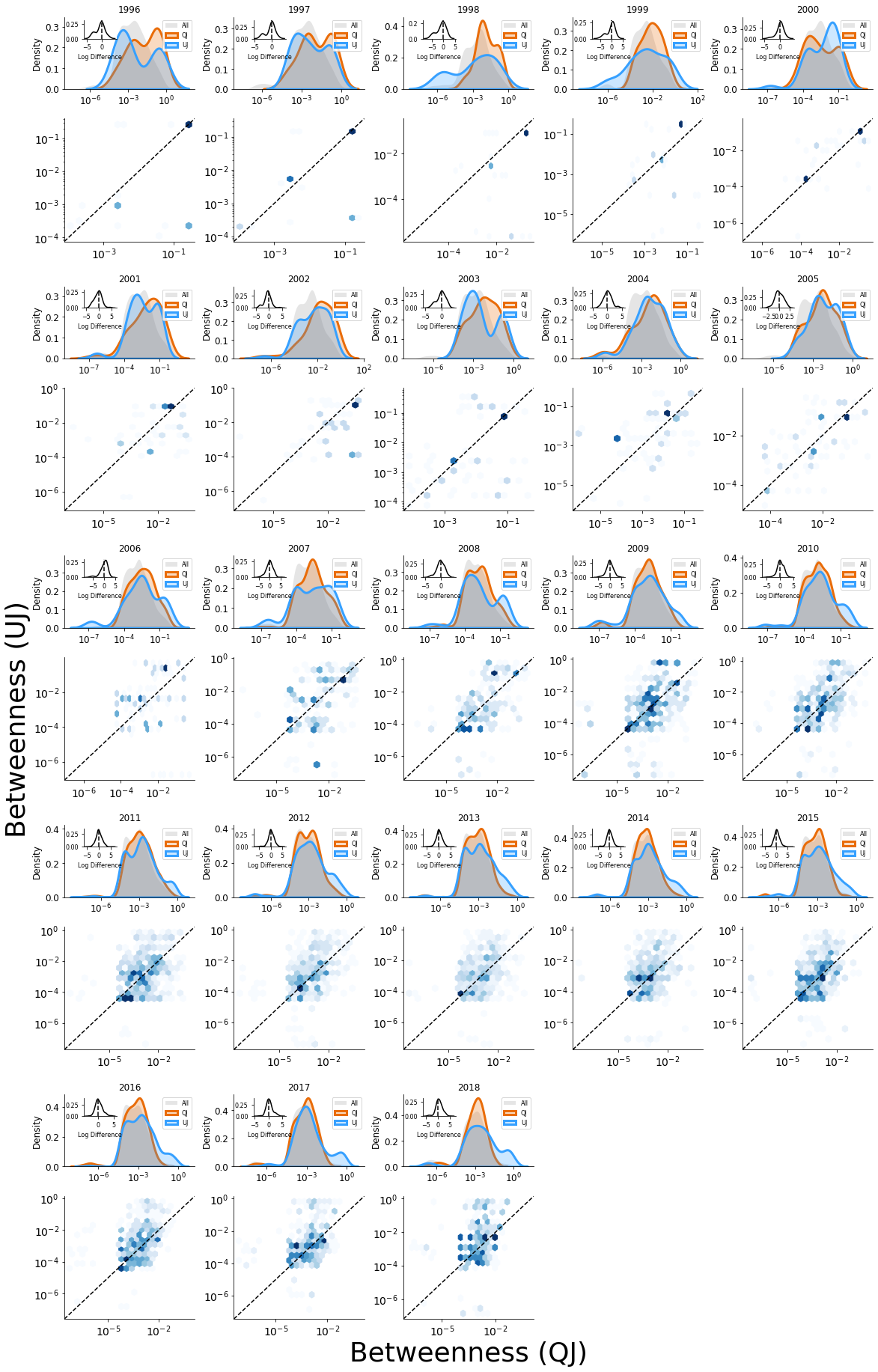}
    \caption{Betweenness centrality computed in the journal citation network with five years of citation time window.}
    \label{fig:sm_BC5y}
\end{figure}

\begin{figure}
    \centering
    \includegraphics[width=0.85\textwidth]{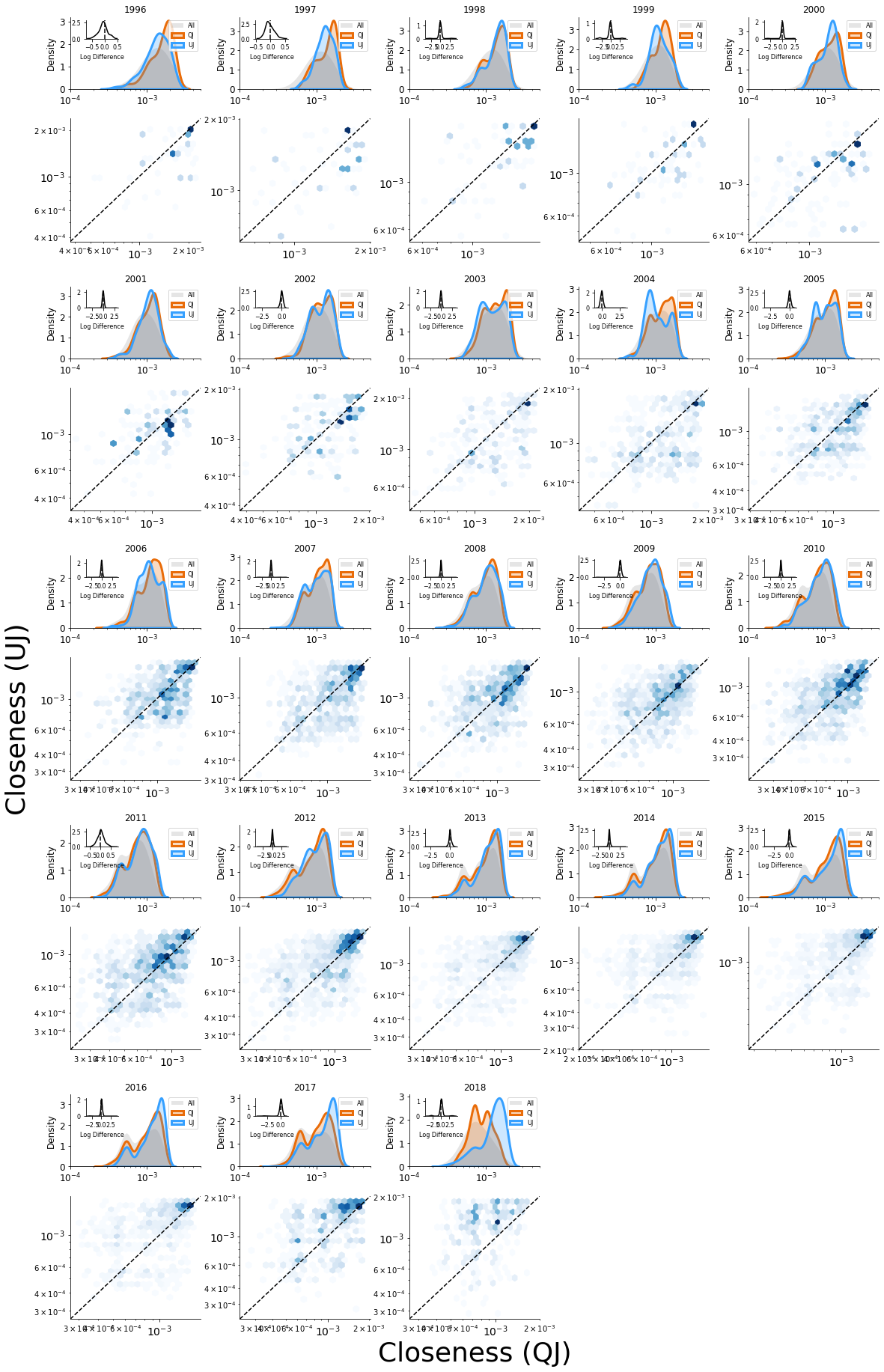}
    \caption{Closeness centrality computed in the journal citation network with five years of citation time window.}
    \label{fig:sm_CC5y}
\end{figure}

\begin{figure}
    \centering
    \includegraphics[width=0.85\textwidth]{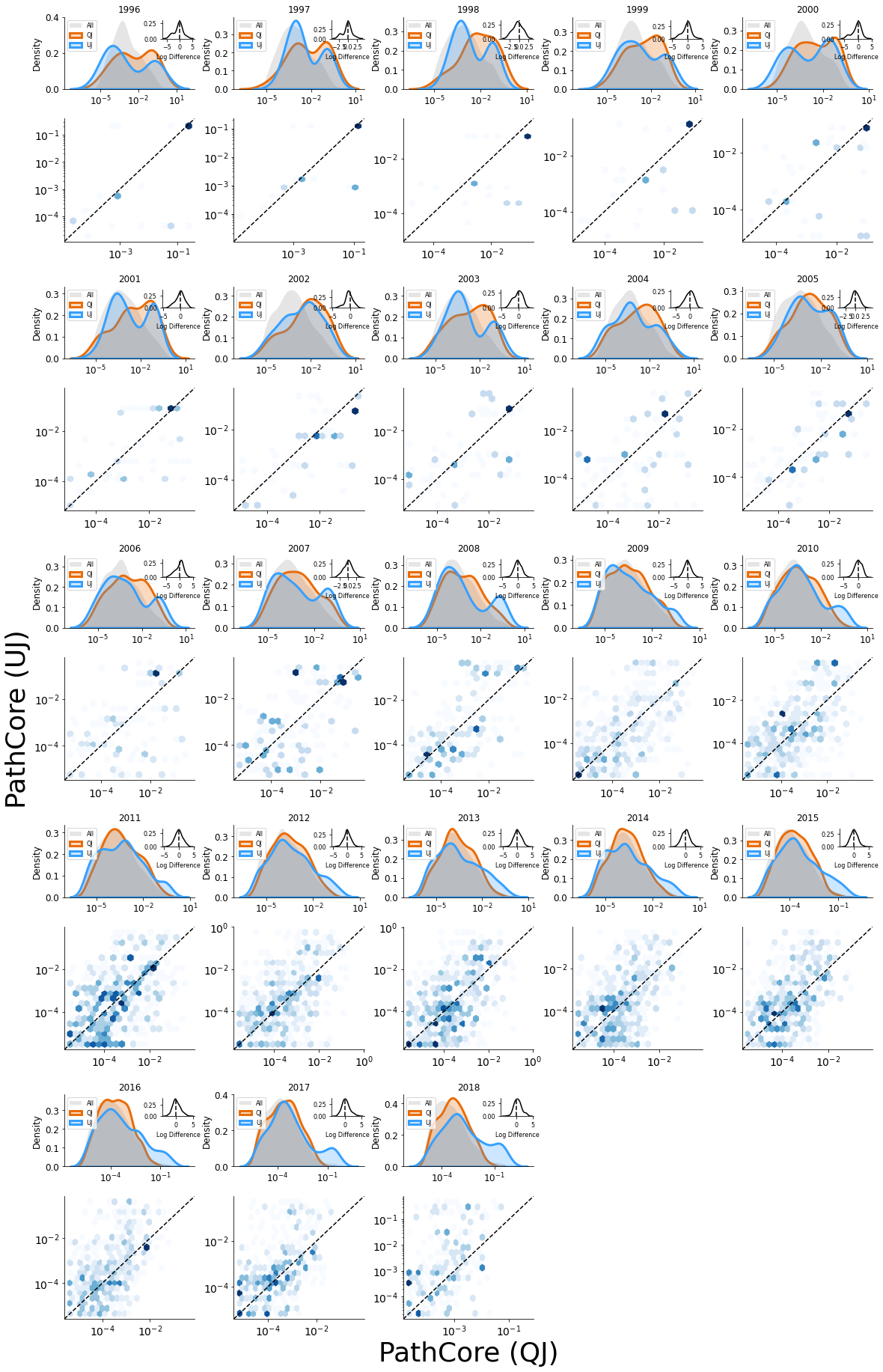}
    \caption{PathCore score computed in the journal citation network with five years of citation time window.}
    \label{fig:sm_PC5y}
\end{figure}

\begin{figure}
    \centering
    \includegraphics[width=0.85\textwidth]{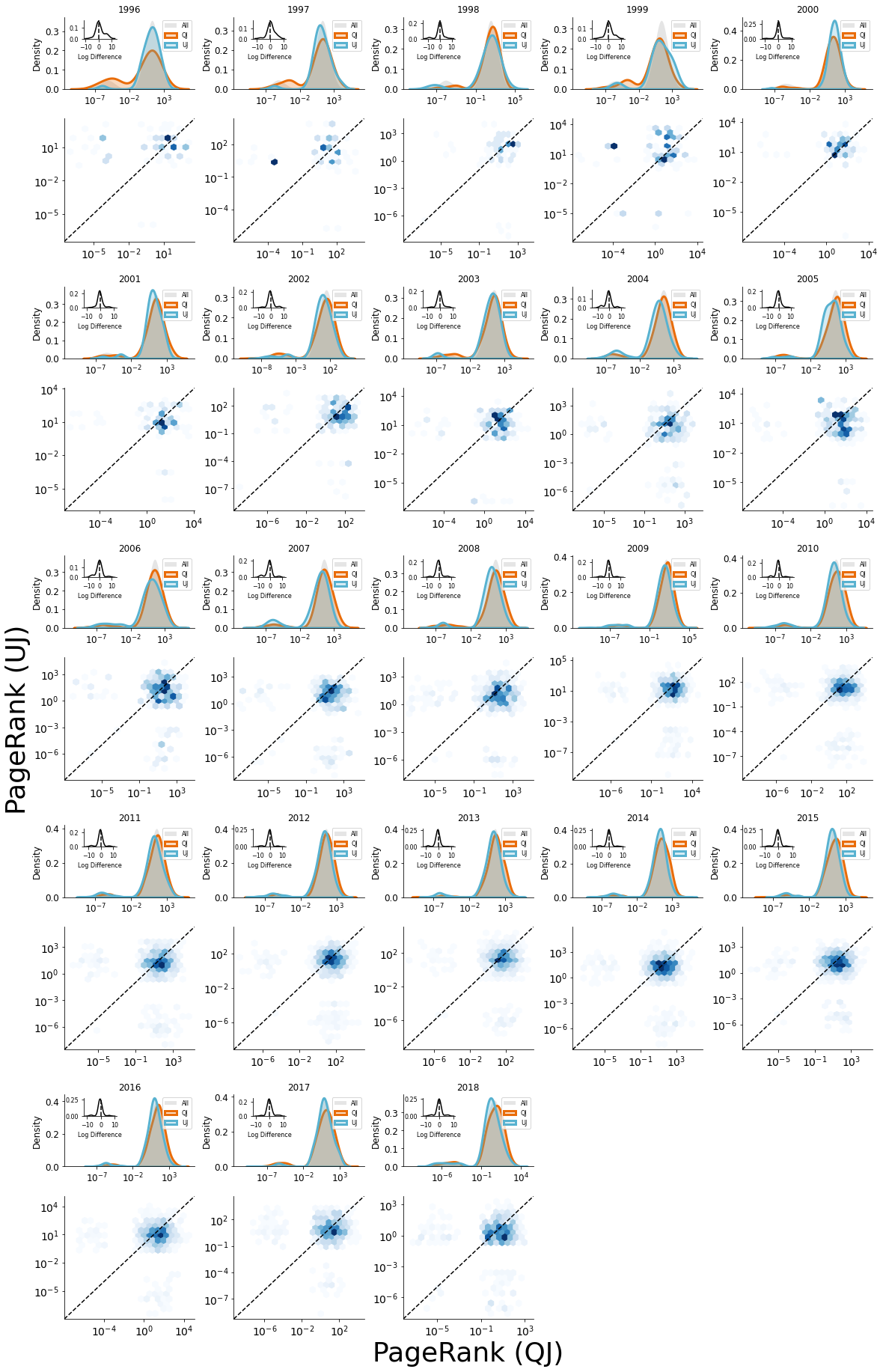}
    \caption{PageRank centrality computed in the journal citation network with five years of citation time window.}
    \label{fig:sm_PR5y}
\end{figure}

\begin{figure}
    \centering
    \includegraphics[width=0.85\textwidth]{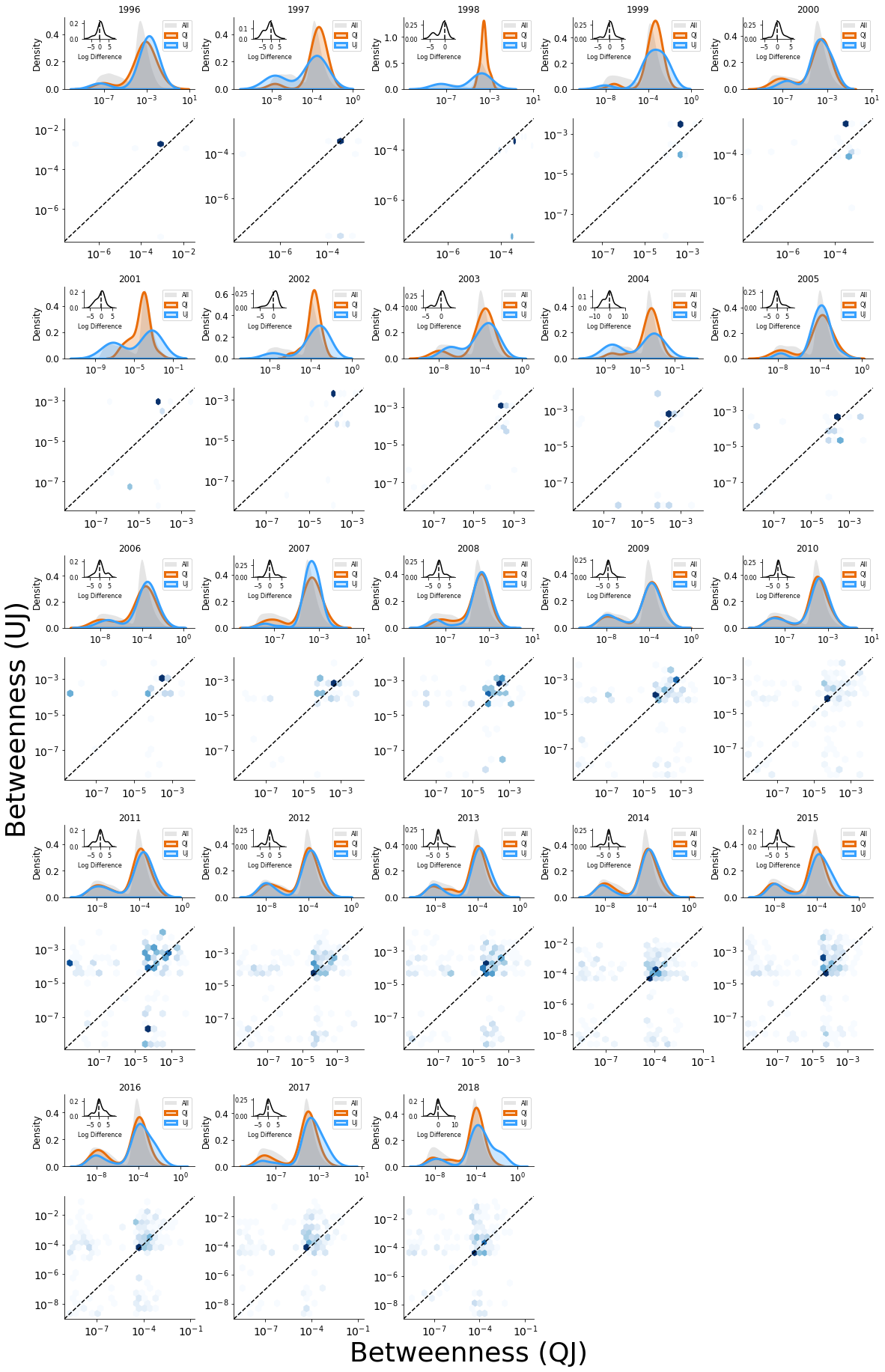}
    \caption{Betweenness centrality computed in the journal citation network with two years of reference time window.}
    \label{fig:sm_BC2y_ref}
\end{figure}

\begin{figure}
    \centering
    \includegraphics[width=0.85\textwidth]{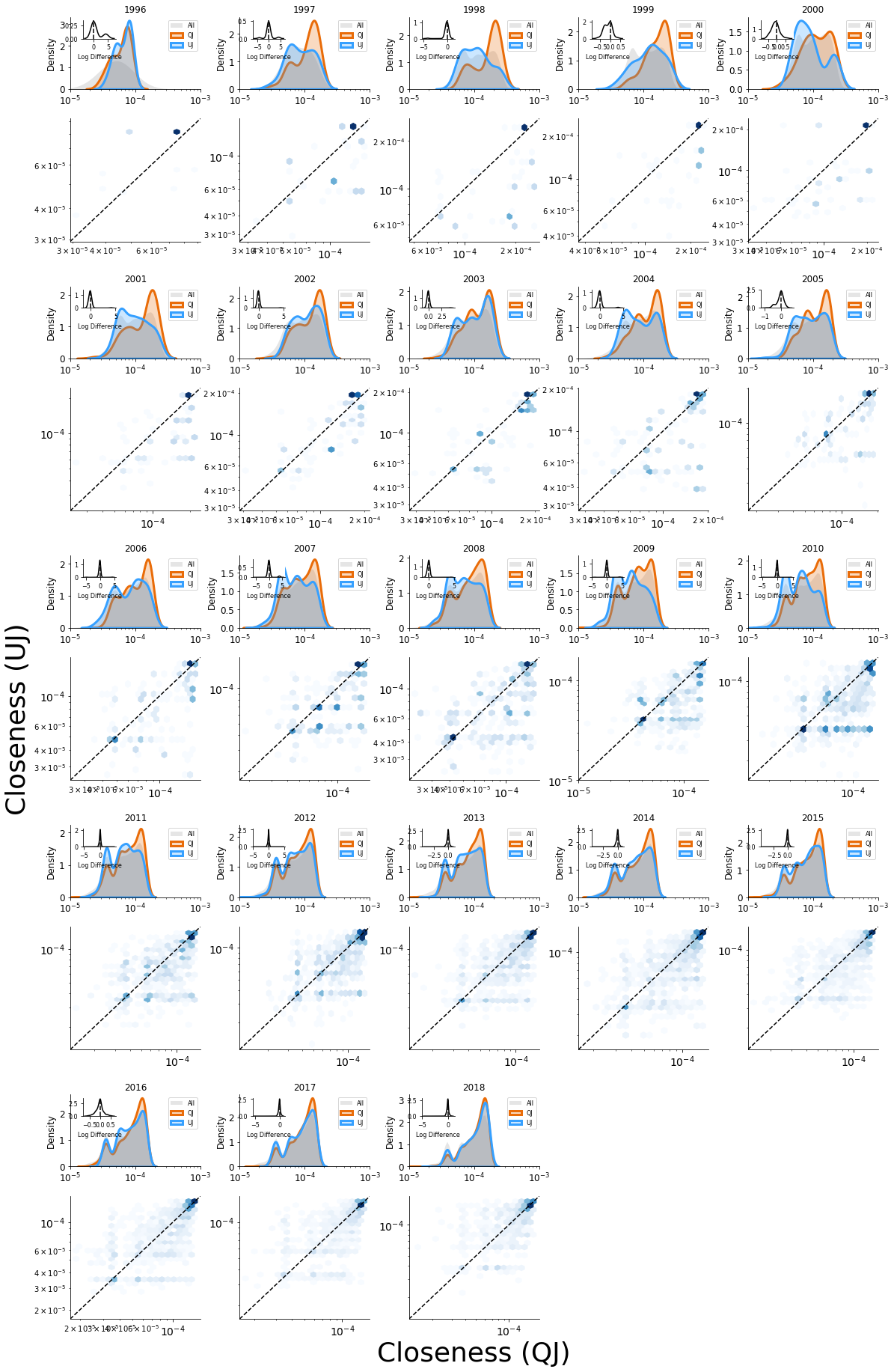}
    \caption{Closeness centrality computed in the journal citation network with two years of reference time window.}
    \label{fig:sm_CC2y_ref}
\end{figure}

\begin{figure}
    \centering
    \includegraphics[width=0.85\textwidth]{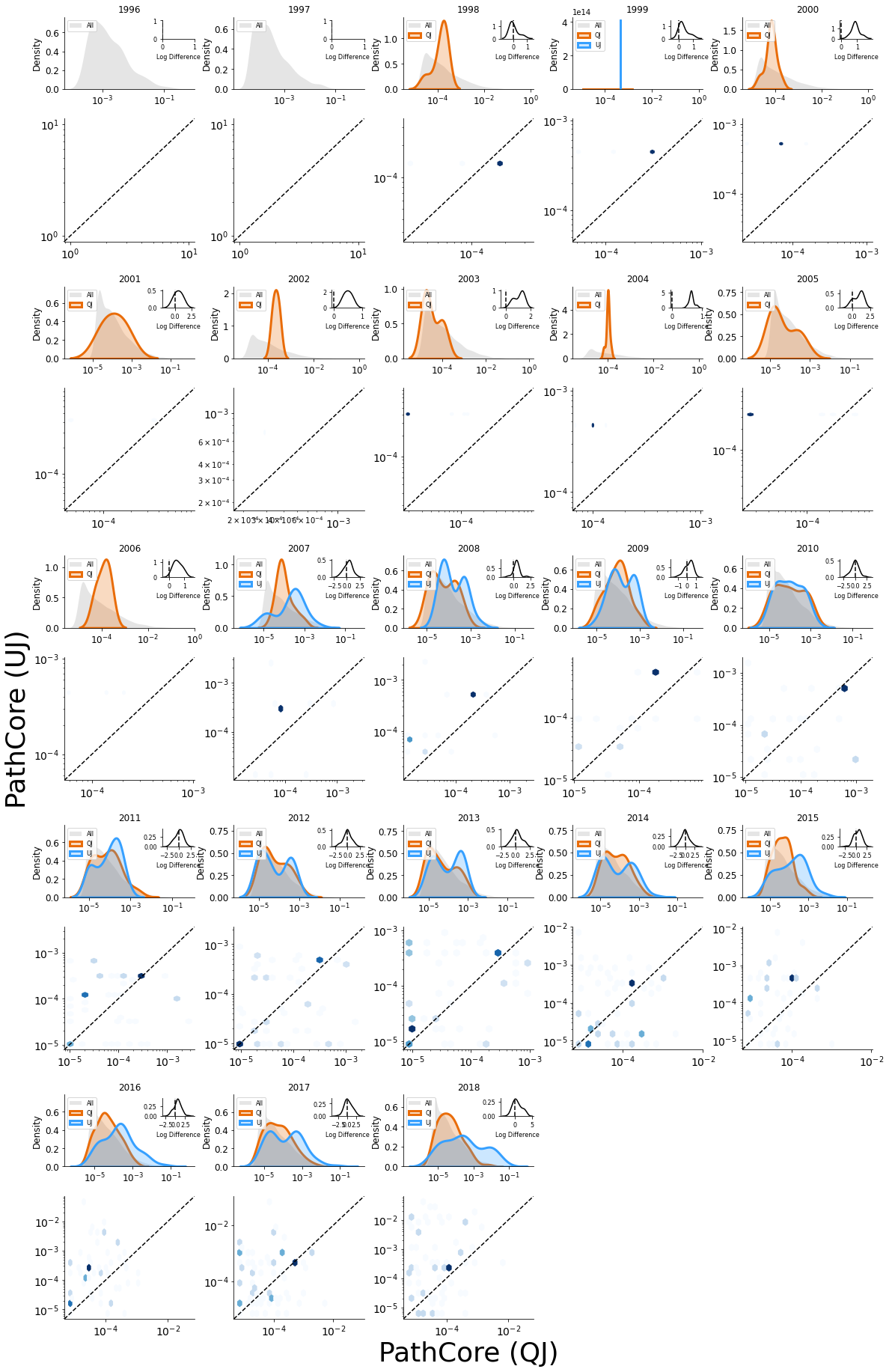}
    \caption{PathCore score computed in the journal citation network with two years of reference time window.}
    \label{fig:sm_PC2y_ref}
\end{figure}

\begin{figure}
    \centering
    \includegraphics[width=0.85\textwidth]{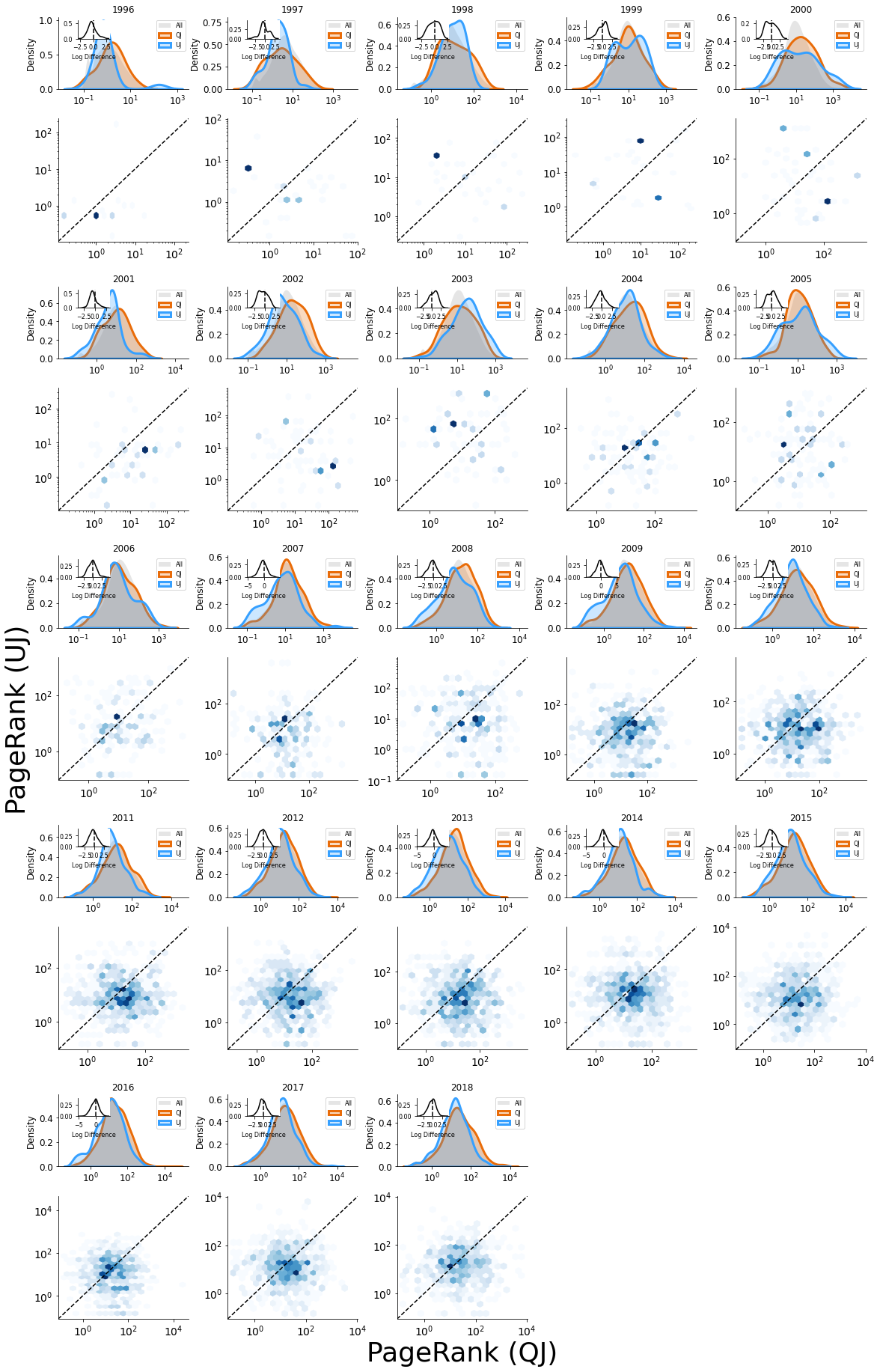}
    \caption{PageRank centrality computed in the journal citation network with two years of reference time window.}
    \label{fig:sm_PR2y_ref}
\end{figure}

\begin{figure}
    \centering
    \includegraphics[width=0.85\textwidth]{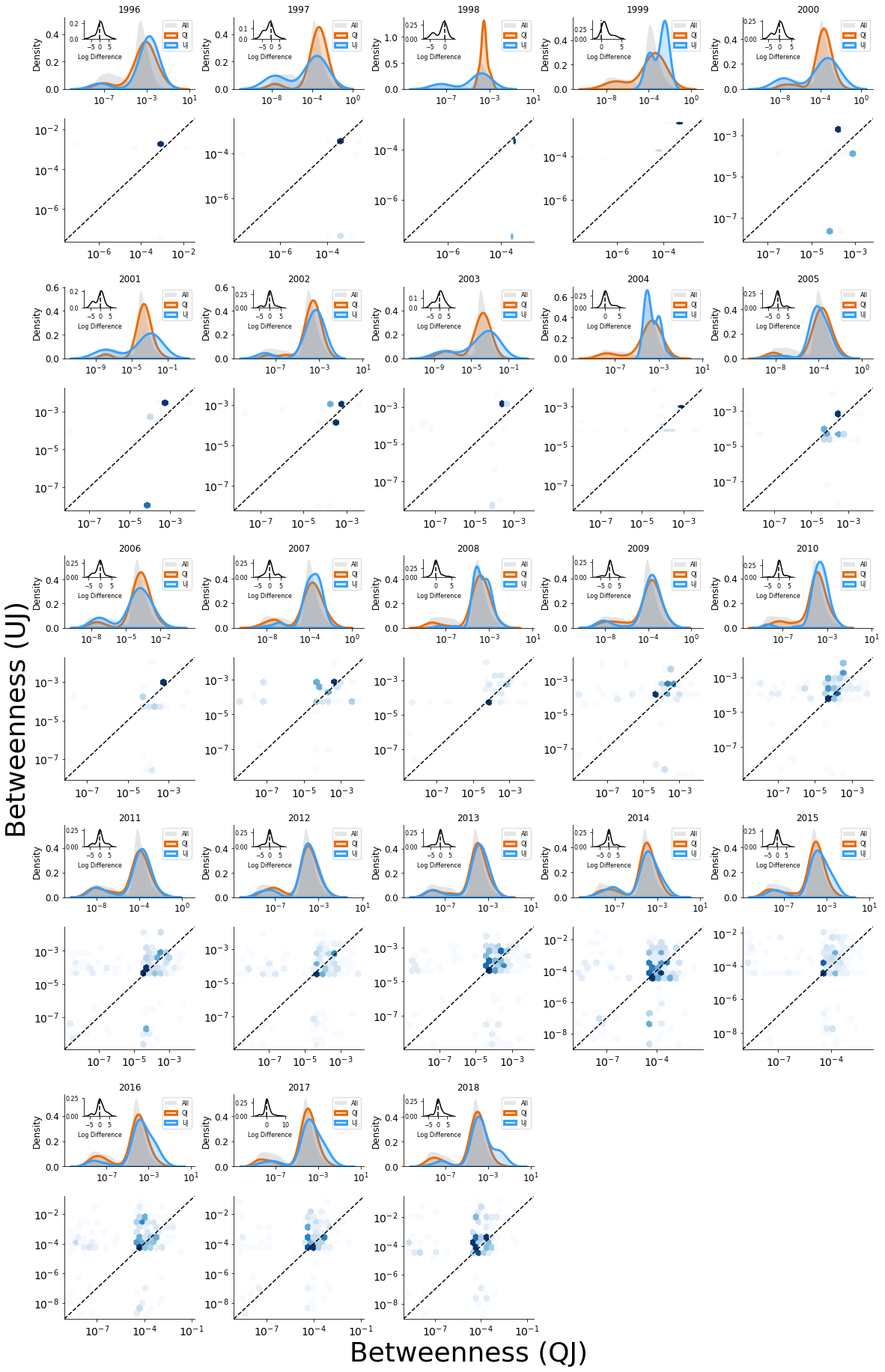}
    \caption{Betweenness centrality computed in the journal citation network with five years of reference time window.}
    \label{fig:sm_BC5y_ref}
\end{figure}

\begin{figure}
    \centering
    \includegraphics[width=0.85\textwidth]{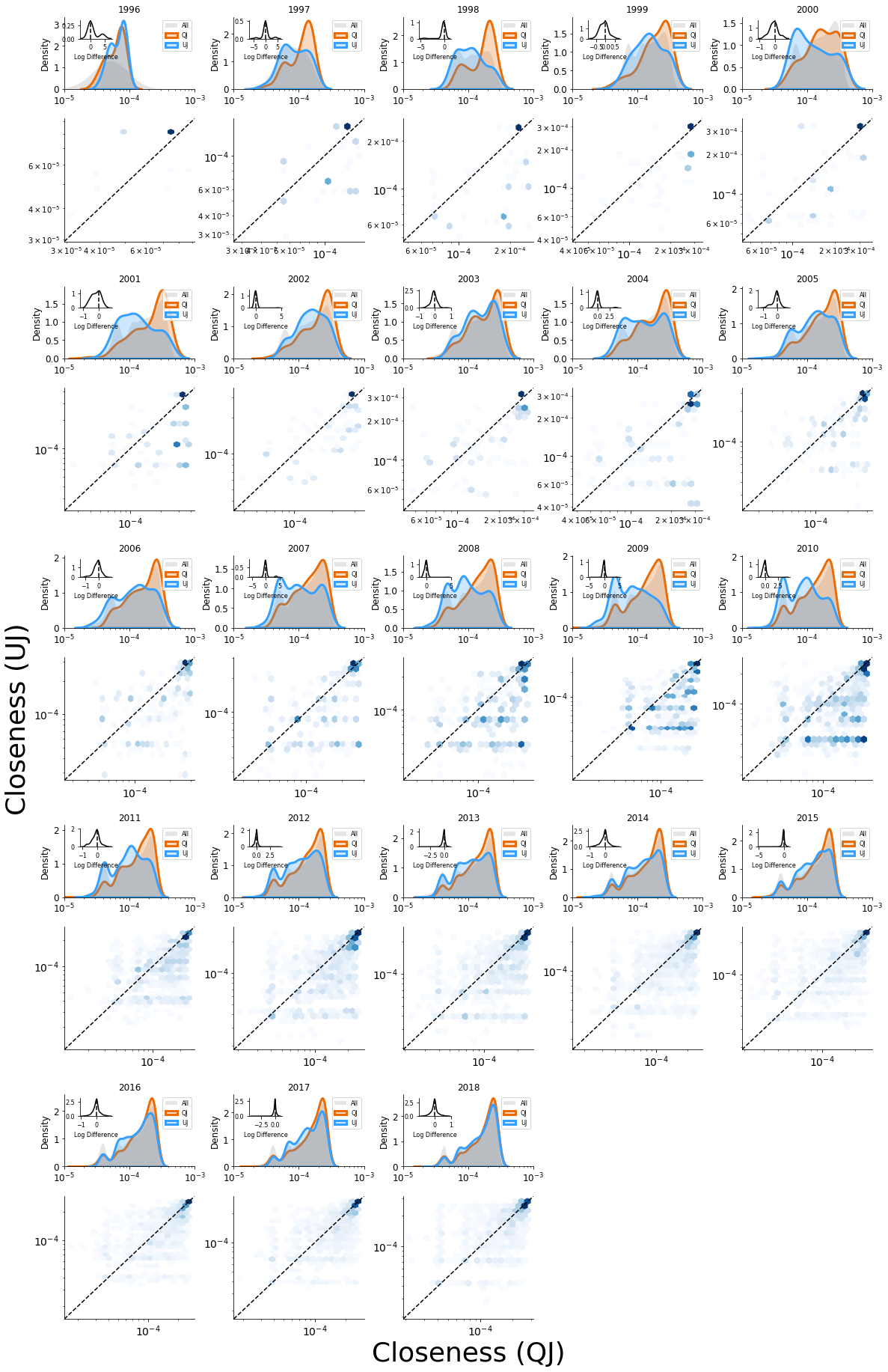}
    \caption{Closeness centrality computed in the journal citation network with five years of reference time window.}
    \label{fig:sm_CC5y_ref}
\end{figure}

\begin{figure}
    \centering
    \includegraphics[width=0.85\textwidth]{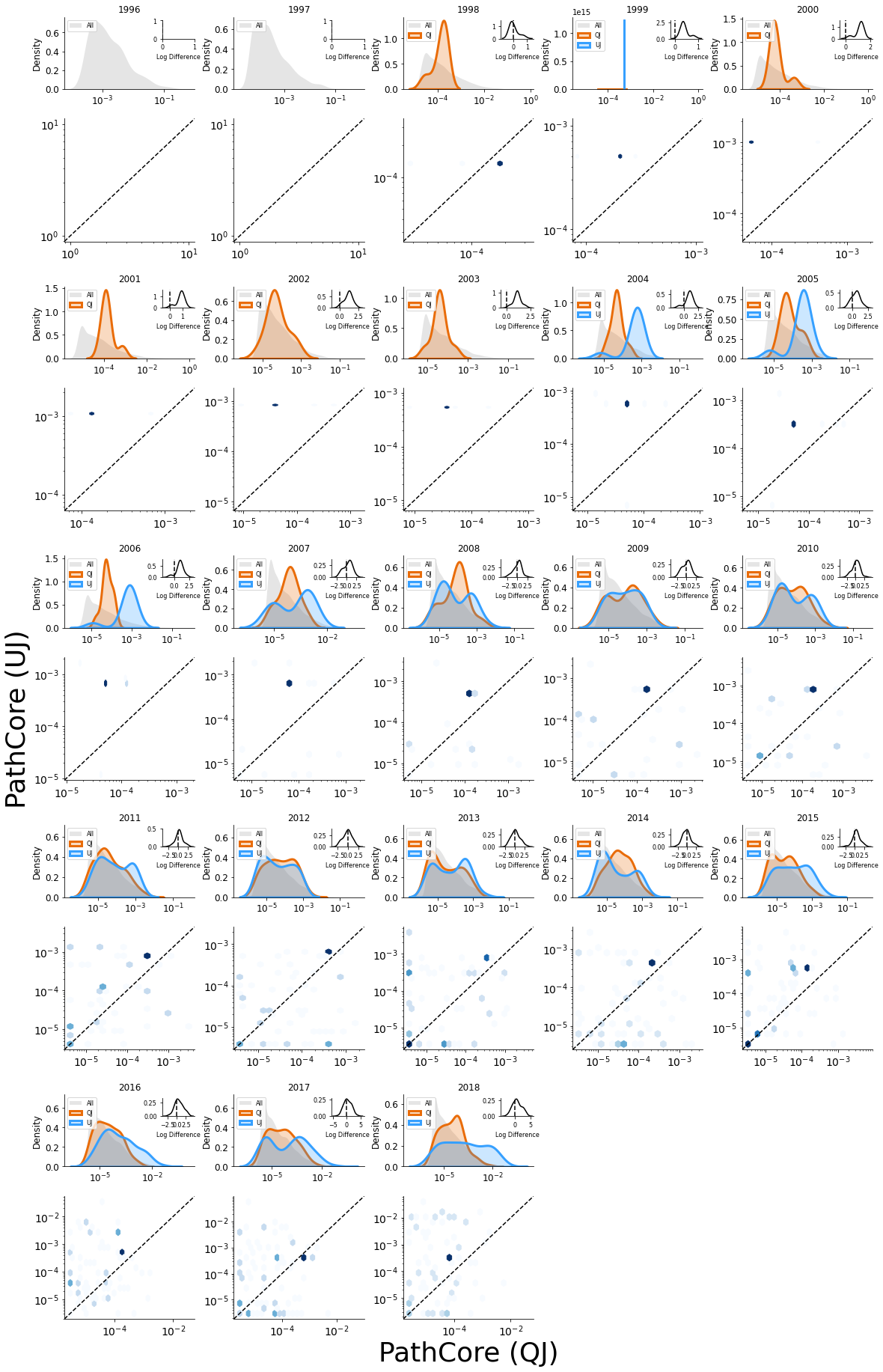}
    \caption{PathCore score computed in the journal citation network with five years of reference time window.}
    \label{fig:sm_PC5y_ref}
\end{figure}

\begin{figure}
    \centering
    \includegraphics[width=0.85\textwidth]{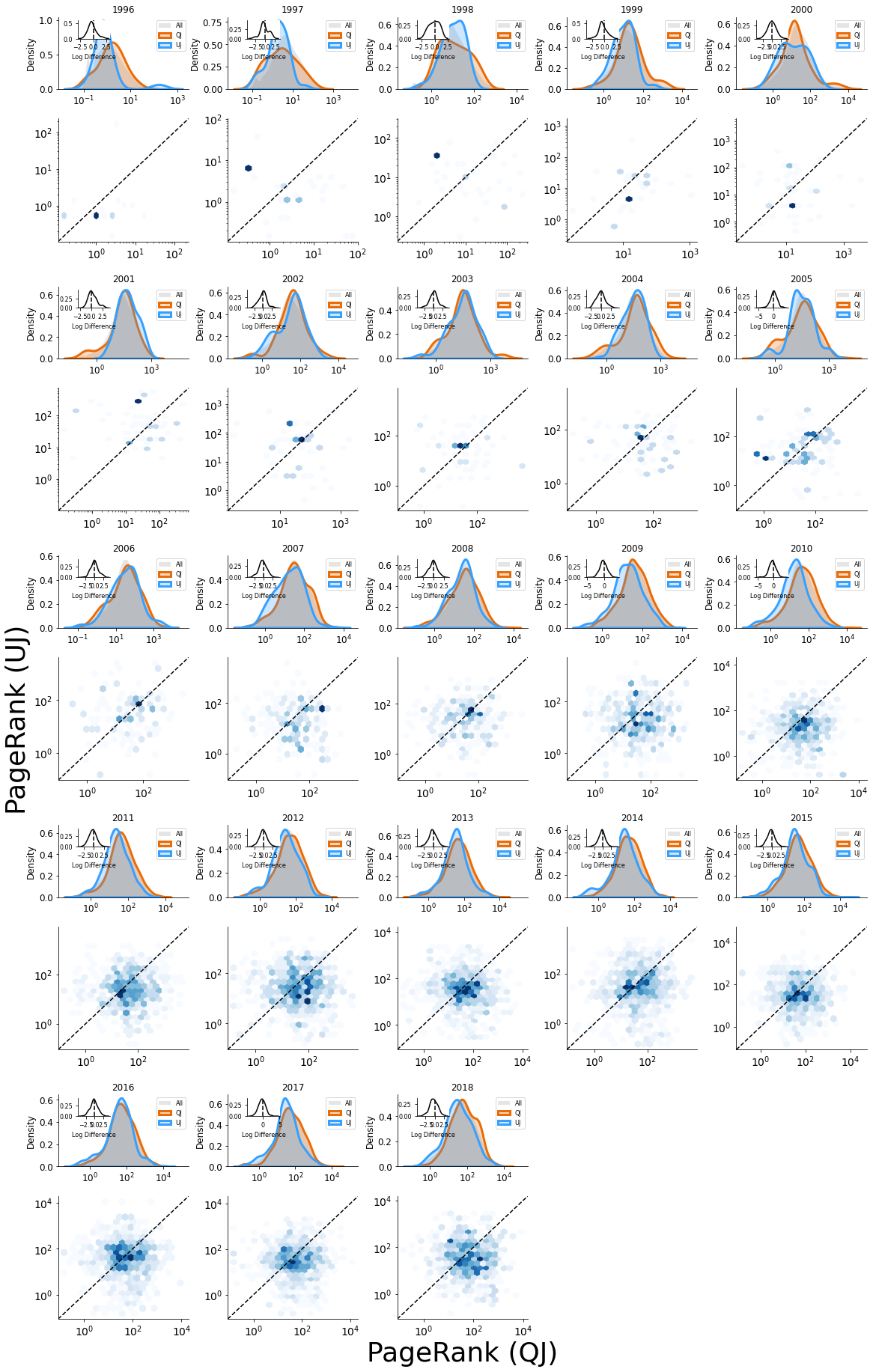}
    \caption{PageRank centrality computed in the journal citation network with five years of reference time window.}
    \label{fig:sm_PR5y_ref}
\end{figure}

In this section, we demonstrate the robustness of the centrality results for network construction. Several types of journal citation networks can be constructed by selecting citations as a link. For instance, one may choose between citations or references of publications and adjust the time window of the citations (references). In addition to the network presented in the main text, we tested four types of networks composed of: i) two-years citation data since publication (Figs.~\ref{fig:sm_BC2y} -- \ref{fig:sm_PR2y}), ii) five-years citation data since publication (Figs.~\ref{fig:sm_BC5y} -- \ref{fig:sm_PR5y}), iii) two-years reference data before publication (Figs.~\ref{fig:sm_BC2y_ref} -- \ref{fig:sm_PR2y_ref}), and iv) five-years reference data before publication (Figs.~\ref{fig:sm_BC5y_ref} -- \ref{fig:sm_PR5y_ref}). Because we focus on the difference between QJs and UJs, we inspected the overall trend rather than the individual value. For instance, Figs.~\ref{fig:sm_BC2y},~\ref{fig:sm_BC5y},~\ref{fig:sm_BC2y_ref}, and~\ref{fig:sm_BC5y_ref} display the BC of the QJs and UJs for the four types of networks. The results indicate that the UJs had a higher BC than the QJs irrespective of the network type. Similarly, other centralities also exhibited robust patterns. Therefore, we concluded that the centrality results are not sensitive to network construction.

\begin{figure}
    \centering
    \includegraphics[width=\textwidth]{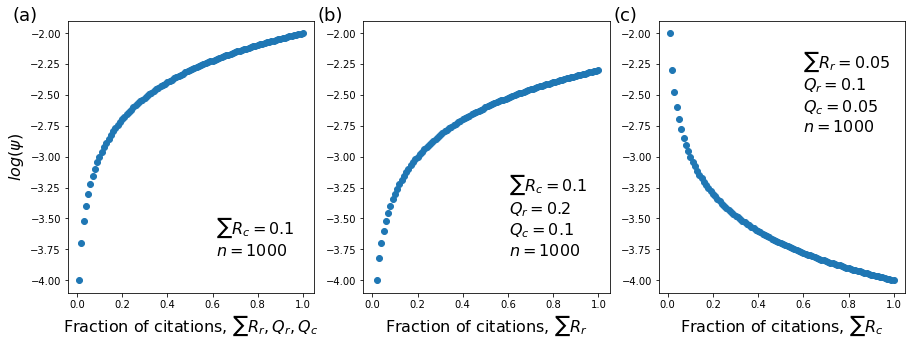}
    \caption{Trend of the publication solidarity index ($\psi$) using synthetic citation counts. (a) A journal gives a considerably large number of citations to its publisher ($\sum R_r = Q_r = Q_c$). (b) A journal gives citations to its publishers but it is not the prevailing amount for its publisher. Thus, the quantity of journal citations ($\sum R_r$) varies, but self-citation rates ($Q_r$ and $Q_c$) remain constant. (c) A journal exclusively receives citations from the publisher.}
    \label{fig:sm_psi_validation}
\end{figure}

\end{document}